\documentclass[pre,twocolumn,showpacs,square,amssymb,amsmath,nobibnotes]{revtex4-1}
\usepackage{bm}
\usepackage{times}
\usepackage{graphicx}
\usepackage[usenames,dvipsnames,svgnames,table]{xcolor}
\usepackage{hyperref}
\usepackage{physics}
\hypersetup{colorlinks=true,  linkcolor=NavyBlue, citecolor=PineGreen,urlcolor=cyan}
\usepackage{physics}
\usepackage{comment}
\usepackage{xcolor}

\hypersetup{colorlinks=true, linkcolor=NavyBlue, citecolor=PineGreen,urlcolor=cyan}

\newcommand{\red}[1]{{\color{red}#1}}

\newcommand{\nn}{\nonumber}
\newcommand{\be}{\begin{equation}}
\newcommand{\ee}{\end{equation}}
\DeclareMathOperator{\sign}{sign}
\newcommand{\moy}[1]{\ensuremath{\langle #1 \rangle}}
\newcommand{\Prob}{{\rm Prob}}

\begin{document}

\global\long\def\so#1{\red{\sout{#1}}}
\global\long\def\l{\lambda}%
\global\long\def\ints{\mathbb{Z}}%
\global\long\def\nat{\mathbb{N}}%
\global\long\def\re{\mathbb{R}}%
\global\long\def\com{\mathbb{C}}%
\global\long\def\dff{\triangleq}%
\global\long\def\df{\coloneqq}%
\global\long\def\del{\nabla}%
\global\long\def\cross{\times}%
\global\long\def\der#1#2{\frac{d#1}{d#2}}%
\global\long\def\bra#1{\left\langle #1\right|}%
\global\long\def\ket#1{\left|#1\right\rangle }%
\global\long\def\braket#1#2{\left\langle #1|#2\right\rangle }%
\global\long\def\ketbra#1#2{\left|#1\right\rangle \left\langle #2\right|}%
\global\long\def\paulix{\begin{pmatrix}0  &  1\\
 1  &  0 
\end{pmatrix}}%
\global\long\def\pauliy{\begin{pmatrix}0  &  -i\\
 i  &  0 
\end{pmatrix}}%
\global\long\def\sinc{\mbox{sinc}}%
\global\long\def\ft{\mathcal{F}}%
\global\long\def\dg{\dagger}%
\global\long\def\bs#1{\boldsymbol{#1}}%
\global\long\def\norm#1{\left\Vert #1\right\Vert }%
\global\long\def\H{\mathcal{H}}%
\global\long\def\tens{\varotimes}%
\global\long\def\rationals{\mathbb{Q}}%
 
\global\long\def\tri{\triangle}%
\global\long\def\lap{\triangle}%
\global\long\def\e{\varepsilon}%
\global\long\def\broket#1#2#3{\bra{#1}#2\ket{#3}}%
\global\long\def\dv{\del\cdot}%
\global\long\def\eps{\gamma}%
\global\long\def\rot{\vec{\del}\cross}%
\global\long\def\pd#1#2{\frac{\partial#1}{\partial#2}}%
\global\long\def\L{\mathcal{L}}%
\global\long\def\inf{\infty}%
\global\long\def\d{\delta}%
\global\long\def\D{\Delta}%
\global\long\def\r{\rho}%
\global\long\def\hb{\hbar}%
\global\long\def\s{\sigma}%
\global\long\def\t{\tau}%
\global\long\def\O{\Omega}%
\global\long\def\a{\alpha}%
\global\long\def\b{\beta}%
\global\long\def\th{\theta}%
\global\long\def\l{\lambda}%

\global\long\def\Z{\mathcal{Z}}%
\global\long\def\z{\zeta}%
\global\long\def\ord#1{\mathcal{O}\left(#1\right)}%
\global\long\def\ua{\uparrow}%
\global\long\def\da{\downarrow}%
 
\global\long\def\co#1{\left[#1\right)}%
\global\long\def\oc#1{\left(#1\right]}%
\global\long\def\tr{\mbox{tr}}%
\global\long\def\o{\omega}%
\global\long\def\nab{\del}%
\global\long\def\p{\psi}%
\global\long\def\pro{\propto}%
\global\long\def\vf{\varphi}%
\global\long\def\f{\phi}%
\global\long\def\mark#1#2{\underset{#2}{\underbrace{#1}}}%
\global\long\def\markup#1#2{\overset{#2}{\overbrace{#1}}}%
\global\long\def\ra{\rightarrow}%
\global\long\def\cd{\cdot}%
\global\long\def\v#1{\vec{#1}}%
\global\long\def\fd#1#2{\frac{\d#1}{\d#2}}%
\global\long\def\P{\Psi}%
\global\long\def\dem{\overset{\mbox{!}}{=}}%
\global\long\def\Lam{\Lambda}%
 
\global\long\def\m{\mu}%
\global\long\def\n{\nu}%

\global\long\def\ul#1{\underline{#1}}%
\global\long\def\at#1#2{\biggl|_{#1}^{#2}}%
\global\long\def\lra{\leftrightarrow}%
\global\long\def\var{\mbox{var}}%
\global\long\def\E{\mathcal{E}}%
\global\long\def\Op#1#2#3#4#5{#1_{#4#5}^{#2#3}}%
\global\long\def\up#1#2{\overset{#2}{#1}}%
\global\long\def\down#1#2{\underset{#2}{#1}}%
\global\long\def\lb{\biggl[}%
\global\long\def\rb{\biggl]}%
\global\long\def\RG{\mathfrak{R}_{b}}%
\global\long\def\g{\gamma}%
\global\long\def\Ra{\Rightarrow}%
\global\long\def\x{\xi}%
\global\long\def\c{\chi}%
\global\long\def\res{\mbox{Res}}%
\global\long\def\dif{\mathbf{d}}%
\global\long\def\dd{\mathbf{d}}%
\global\long\def\grad{\vec{\del}}%

\global\long\def\mat#1#2#3#4{\left(\begin{array}{cc}
 #1  &  #2\\
 #3  &  #4 
\end{array}\right)}%
\global\long\def\col#1#2{\left(\begin{array}{c}
 #1\\
 #2 
\end{array}\right)}%
\global\long\def\sl#1{\cancel{#1}}%
\global\long\def\row#1#2{\left(\begin{array}{cc}
 #1  &  ,#2\end{array}\right)}%
\global\long\def\roww#1#2#3{\left(\begin{array}{ccc}
 #1  &  ,#2  &  ,#3\end{array}\right)}%
\global\long\def\rowww#1#2#3#4{\left(\begin{array}{cccc}
 #1  &  ,#2  &  ,#3  &  ,#4\end{array}\right)}%
\global\long\def\matt#1#2#3#4#5#6#7#8#9{\left(\begin{array}{ccc}
 #1  &  #2  &  #3\\
 #4  &  #5  &  #6\\
 #7  &  #8  &  #9 
\end{array}\right)}%
\global\long\def\su{\uparrow}%
\global\long\def\sd{\downarrow}%
\global\long\def\coll#1#2#3{\left(\begin{array}{c}
 #1\\
 #2\\
 #3 
\end{array}\right)}%
\global\long\def\h#1{\hat{#1}}%
\global\long\def\colll#1#2#3#4{\left(\begin{array}{c}
 #1\\
 #2\\
 #3\\
 #4 
\end{array}\right)}%
\global\long\def\check{\checked}%
\global\long\def\v#1{\vec{#1}}%
\global\long\def\S{\Sigma}%
\global\long\def\F{\Phi}%
\global\long\def\M{\mathcal{M}}%
\global\long\def\G{\Gamma}%
\global\long\def\im{\mbox{Im}}%
\global\long\def\til#1{\tilde{#1}}%
\global\long\def\kb{k_{B}}%
\global\long\def\k{\kappa}%
\global\long\def\ph{\phi}%
\global\long\def\el{\ell}%
\global\long\def\en{\mathcal{N}}%
\global\long\def\asy{\cong}%
\global\long\def\sbl{\biggl[}%
\global\long\def\sbr{\biggl]}%
\global\long\def\cbl{\biggl\{}%
\global\long\def\cbr{\biggl\}}%
\global\long\def\hg#1#2{\mbox{ }_{#1}F_{#2}}%
\global\long\def\J{\mathcal{J}}%
\global\long\def\diag#1{\mbox{diag}\left[#1\right]}%
\global\long\def\sign#1{\mbox{sgn}\left[#1\right]}%
\global\long\def\T{\th}%
\global\long\def\rp{\reals^{+}}%

\title{}

\title{Extreme value statistics for branching run-and-tumble particles}
\author{Bertrand Lacroix-A-Chez-Toine}
\affiliation{Department of Physics of Complex Systems, Weizmann Institute of Science,
Rehovot 7610001, Israel}
\author{Asaf Miron}
\affiliation{Department of Physics of Complex Systems, Weizmann Institute of Science,
Rehovot 7610001, Israel}

\begin{abstract}

The extreme value statistics of active matter offer significant insight into their unique properties. A phase transition has recently been reported in a model of branching run-and-tumble particles, describing the spatial spreading of an evolving colony of active matter in one-dimension. In a "persistent" phase, the particles form macroscopic robust clusters that ballistically propagate as a whole while in an "intermittent" phase, particles are isolated instead. We focus our study on the fluctuations of the rightmost position $x_{\max}(t)$ reached by time $t$ for this model. At long time, as the colony  
progressively invades the unexplored region, the cumulative probability of $x_{\max}(t)$ is described by a travelling front. The transition has a remarkable impact on this 
front.
In the intermittent phase it is qualitatively similar to the 
front satisfying the Fisher-KPP equation, which famously describes the extreme value statistics of the \textit{non-active} branching Brownian motion. A dramatically different behaviour appears in the persistent phase, where activity imparts the front with unexpected and unusual features which we compute exactly. 
\end{abstract}

\maketitle

\section{Introduction}

Systems whose constituents
posses an innate ability to convert energy available in their environment into directed
motion are commonly referred-to as active matter. In the context of biological systems, it encapsulates
a broad range of living phenomena, ranging from macroscopic multi-cellular
organisms like animals, down to single-cell organisms, such as
bacteria and their sub-cellular components \cite{sanchez2012spontaneous,RevModPhys.85.1143,doi:10.1146/annurev-conmatphys-031214-014710,Ramaswamy2019}. 

A fundamental characteristic of active matter is the breaking of detailed balance associated with the generation of directed motion. 
This leads to a wealth of unique features observed both on the collective level, such as self-organization, flocking and motility-induced phase separation \cite{vicsek1995novel,Czir_k_1997,TONER2005170,Buhl1402,Bertin_2009,doi:10.1146/annurev-conmatphys-031214-014710,PhysRevLett.116.218101,PhysRevLett.75.2899,doi:10.1080/000187300405228}, as well as for individual particles, which may exhibit a non-Gibbsian equilibrium distribution \cite{PhysRevE.99.032132,Malakar_2018}. 
Many of these interesting phenomena are 
known to be closely related to active matter's ability to persistently maintain directed motion over 
extended periods of time. It is therefore not surprising that recent years have seen a resurgence of interest in persistent random walk models \cite{Furth1920244,taylor1922diffusion,Goldstein1951129,stadje1987exact,weiss2002some,masoliver2017continuous}.
One popular example is the "run-and-tumble" particle (RTP), which has been shown to describe the motility of various bacteria, such as E. coli \cite{PhysRevLett.100.218103,le2019noncrossing,PhysRevLett.116.218101,berg2008coli,PhysRevLett.123.260602,PhysRevLett.123.250603,PhysRevLett.123.238004,PhysRevLett.122.068002,Cates_2012,das2019gap}. In this model, particles have a fixed velocity $v_0$ whose direction changes in "tumbling" events that occur at a constant rate $\gamma$. In contrast to Brownian motion, where particles are subjected to random forces preventing any persistence, the RTP displays a finite persistence length $v_0/\gamma$.

Another important facet of living active systems is their ability to multiply and form extended colonies. Several frameworks have been used to model the spatial spreading of bacterial colonies. One well-studied macroscopic approach adopts an effective description in terms of reaction-diffusion equations \cite{GOLDING1998510,MIMURA2000283,MATSUSHITA1990498,PhysRevLett.108.088102,wakita1994experimental,wakano2004phase}. A different, more microscopic, approach instead models bacteria 
as branching Brownian motion (BBM) \cite{arguin2013extremal,PhysRevLett.112.210602,PhysRevLett.77.4780,harris1964theory,ramola2015spatial,ramola2015branching,brunet2011branching,PhysRevLett.122.020602,dumonteil2013spatial}, where each Brownian particle may branch into two identical particles or die. In 1D
, the spreading of BBM is 
characterised by the evolution of the maximal position $x_{\max}(t)$ reached up to time $t$ by any 
of the particles in the colony.
Branching processes are one of a handful of examples of correlated random processes in which exact results may be derived for the extreme value statistics (EVS), and are typically described by a travelling front (TF) \cite{majumdar2003extreme}. However, there are very few examples where the functional form of the front can be described analytically \cite{ablowitz1979explicit}.
In the context of BBM, seminal works by McKean \cite{doi:10.1002/cpa.3160280302}
and Bramson \cite{bramson1983convergence} have shown that the cumulative distribution function (CDF) of $x_{\max}\left(t\right)$ is given at large time by the travelling front satisfying 
the well-known Fisher-KPP equation \cite{kolmogorov1937investigation,doi:10.1111/j.1469-1809.1937.tb02153.x}. 


In this article we investigate the EVS of an evolving colony of active matter by revisiting the 1D branching run-and-tumble particle model considered in \cite{HORSTHEMKE1999285,demaerel2019asymptotic}. A phase transition has been uncovered in the velocity of the front describing the EVS of the process by studying the evolution of the density of particles \cite{demaerel2019asymptotic}.
However, the physical origin of this phase transition and its impact on the overall shape of the travelling front have not been analysed. Previous numerical 
studies of reaction-diffusion models for bacterial spreading 
have established that such 
a phase transition 
has a dramatic impact on the travelling 
front's shape \cite{wakano2004phase} but no clear analytical description has been given to this phenomenon.
We fill this gap 
by obtaining exact analytical results 
demonstrating the impact of the phase transition on the shape of the distribution of the rightmost position $x_{\max}(t)$. The "intermittent" phase features a front similar to the one arising in the \textit{non-active} branching Brownian motion and is solution of the Fisher-KPP equation. In particular, in this phase the front has exponential tails on both ends and its position exhibits
the 
famous universal logarithmic correction with respect to ballistic spreading \cite{ebert2000front}.
In the "persistent" phase we obtain a complete and exact analytical description for the front's functional form and its position. In contrast to the intermittent phase, we show that the front has a finite edge beyond which it vanishes exactly.


The paper is organised as follows. 
In section \ref{sec_mod} we present the 
model 
and explain the origin of the phase 
transition.
In section \ref{sec_res} we detail our main results. In section \ref{sec_eq}, we derive the main equations describing the cumulative probability of the maximum and its associated travelling front.
In section \ref{sec_int_phase}, we describe the properties of the TF in the "intermittent phase". We first obtain its 
velocity and the 
logarithmic correction to 
its position with respect to ballistic spreading, after which we proceed to analyze the TF's asymptotic behaviour and its relation to the branching Brownian motion.
In section \ref{sec_per_phase}, we describe the TF in the "persistent phase". We first show that it presents a finite edge beyond which it vanishes exactly. We then obtain an exact analytical expression for the functional form of the TF and study its asymptotic behaviour.
We then compute the moments of the distribution and consider some limiting cases. In section \ref{sec_ph_trans} we characterise the TF solution at the transition between the two phases. Finally in section \ref{sec_con} we conclude and give some future directions. 



\section{Model}\label{sec_mod}
\subsection{Model and observables}

We consider a 1D 
model of branching RTP. At time $t=0$, a single RTP lies at $x_0=0$ 
with velocity $\sigma v_{0}$, directed either to the right ($\sigma=+$) or to the left ($\sigma=-$). At any later time $t>0$, the system contains $N(t)$ particles 
$k\in\{1,\cdots,N(t)\}$ characterised by 
their position $x_k(t)$ and direction of motion $\sigma_k(t)=\pm$. The system evolves according to the following stochastic rules: 
during a small time interval $dt$ each particle $k$ may, (i) branch into two independent particles with probability $dt$, (ii) die with probability $\delta dt$, (iii) tumble, reversing its direction $\sigma_k(t+dt)=-\sigma_k(t)$, with probability $\gamma dt$ or (iv) move ballistically to the new position $x_k(t+dt)=x_k(t)+\sigma_k(t)v_0 dt$ with 
the complementary probability $1-(1+\delta+\gamma)dt$. 
When a particle branches, it produces an offspring moving in either direction with equal probability. We characterise the process' spreading by considering the evolution of the rightmost position reached by any RTP in the history of the process up to time $t$,

\be
x_{\max}(t)=\displaystyle \max_{\tau\in[0,t]}\{x_{k}(\tau)\}_{k=0}^{N(\tau)}\;.
\ee
We compute the exact large-time 
behaviour of the (complementary) cumulative probability 

\be
Q_{\sigma}(x,t)\equiv\Prob\left[x_{\max}(t)\geq x|x_{0}=0,\sigma\right]\;,\label{Q_max_def}
\ee
that $x_{\max}(t)$ remains bounded to $[x,+\infty)$ for time $t$, given the initial position $x_0=0$ and direction $\sigma$ of the initial RTP. 

One must first distinguish between the case $\d\ge1$, where the number of particles $N(t)$ eventually goes to zero with probability $P(t)=\Prob[N(t)>0]\to 0$, and $\delta<1$, where the process has a finite probability $P(t)\to 1-\delta$ to survive in the long-time limit (See App. \ref{app_num} for details). 
In this article we restrict our discussion to the latter, $0\leq\delta<1$, where the average number of RTPs grows exponentially over time, 
$\moy{N(t)}=e^{(1-\delta)t}$. Of course, 
if the colony does go extinct 
$x_{\max}(t)$ 
retains its value at the time of extinction 
and the probability $Q_{\sigma}(x,t)$ 
reaches a stationary state $Q_{\sigma}^{\rm st}\left(x\right)$. 
Conditioning on the survival or extinction 
of the process, i.e. $N(t)>0$ or $N(t)=0$, the cumulative probability reaches at large time the asymptotic scaling form
\begin{equation}
Q_{\sigma}(x,t)\approx \delta\,Q_{\sigma}^{\rm st}\left(x\right)+(1-\delta)\,F_{\sigma}\left(\zeta=x-m(t) \right)\;.\label{late_t_cond}
\end{equation}
This scaling form is observed clearly on Fig. \ref{Fig_F_int}. The function $F_{\sigma}(\zeta)$ describes the travelling front which arises if the colony does survive up to time $t$. Note that it also describes the cumulative probability of the rightmost position over all particles $k\in\{1,\cdots,N(t)\}$ alive at 
time $t$. The parameter $m(t)$ represents the front's position at time $t$ and is given by,
\be
m(t)=v t+X(t)\; \; {\rm where}\;\;X(t)=o(t)\;.\label{front_pos}
\ee
At leading order, the front propagates ballistically with velocity $v$ and $X(t)$ describes the correction with respect to this ballistic motion at large time. 
Note that the position of the front $m(t)$ is not uniquely defined \cite{brunet2011branching}. A common definition is such that the TF takes a fixed value $F(\zeta=0)=f$ for some $0<f<1$.
The leading correction to the front's position has been extensively studied and has been shown to take the universal value $X(t)\propto (3/2)\ln t$ (up to a multiplying non-universal constant) for a general class of non-linear equations \cite{ebert2000front}. 
Yet sub-leading terms which are either of 
$O(1)$ or that vanish in the large-time limit depend on the precise definition of the front's position and are in general hard to obtain exactly \cite{berestycki2017exact,berestycki2018new}. We focus our analysis in this article on this travelling front solution. 

\begin{figure}
    \centering
    \includegraphics[width=0.475\textwidth]{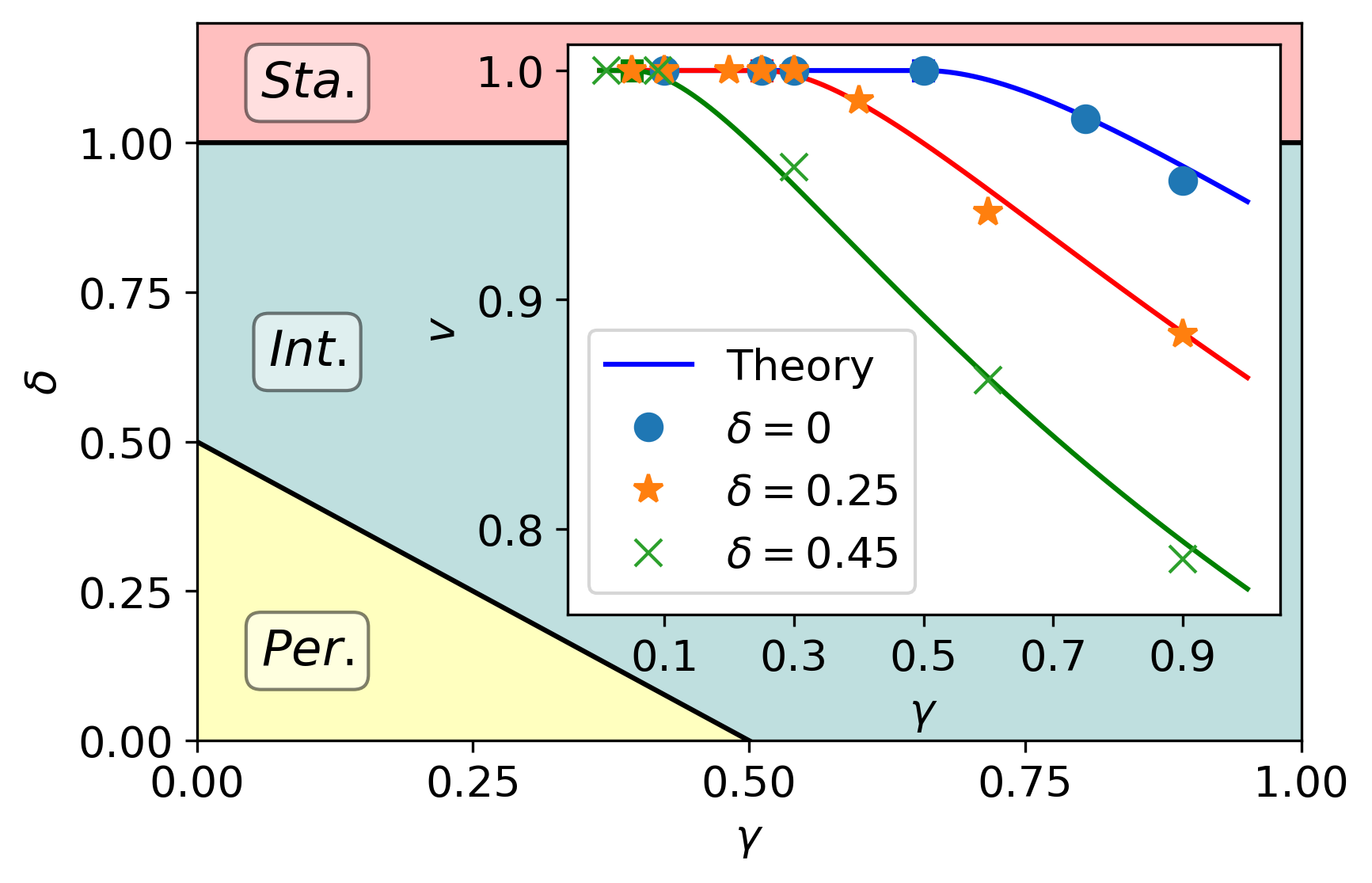}
    \caption{Phase diagram of the model. In the stationary phase, for $\d\ge1$, all particles eventually perish, there is no TF solution and the system reaches a stationary state. For $\d<1$ there are 
    two distinct phases \cite{demaerel2019asymptotic,HORSTHEMKE1999285}: A persistent phase for $\d+\gamma<1/2$ where the velocity is constant and given by $v_0$ and an intermittent phase for $\d+\gamma>1/2$ where the velocity $v_{\rm i}$ is given in Eq. \eqref{v_TW2}. Inset: Comparison between the velocity $v$ obtained from numerical simulation of the stochastic process versus the tumbling rate $\g$ for different dying rates $\d$ and the analytical prediction in Eq. \eqref{v_TW2}. 
    }
    \label{Fig_PD_v}
\end{figure}

\subsection{Phase transition}

This model exhibits a phase transition between a "persistent" phase, in which the particles form ballistically propagating macroscopic clusters, and an "intermittent" phase where the particles are instead isolated. 
We now detail the mechanism at the origin of this transition. 
Any single particle with velocity $\sigma v_{0}$ branches into a pair of independent particles propagating at the same velocity $\sigma v_{0}$ and with the same position at rate 
$1/2$. Yet, once created, this pair separates if either particle tumbles or dies, which occurs with the respective rates $\gamma$ and $\delta$. 
One therefore expects an exponential 
increase in the number of particles in such a ballistically propagating cluster if $\gamma+\delta<1/2$ since then particles moving along the same direction multiply at a faster rate than they are depleted by tumbling and death events.
In the intermittent phase, where 
$\gamma+\delta>1/2$, the number of particles in a cluster instead 
decays exponentially over time
, such that any cluster has a finite life-time after which 
most RTPs evolve separately.

The impact of this phase transition on the TF 
velocity 
was already described for this model in \cite{HORSTHEMKE1999285, demaerel2019asymptotic}  (see also \cite{bouin2015propagation, bouin2014hyperbolic, fedotov1998traveling} for similar phase transitions in distinct models) and is summarised in Fig. \ref{Fig_PD_v}.
In the persistent phase, the rightmost cluster at time $t\gg 1$ contains typically an exponentially large number of particles and as such its evolution is not impacted by the  tumbling or death of individual particles. The evolution of $x_{\max}(t)$ follows that of this cluster and propagates ballistically and persistently with velocity $v=v_0$.
In the intermittent phase, a particle located at $x=x_{\max}(t)$ at time $t$ is typically isolated. After a stochastic time, this particle will eventually die or tumble and
$x_{\max}(t)$ will cease to grow until another right-moving RTP crosses this position. We call this phase intermittent due to the alternate periods of growth and plateau 
of $x_{\max}(t)$. This behaviour still leads to an average ballistic growth, albeit with a smaller velocity $v_{\rm i}\leq v_0$.

\section{Main results}\label{sec_res}

We have described in section \ref{sec_mod} the impact of the phase transition on the velocity of the travelling front (see also \cite{demaerel2019asymptotic,HORSTHEMKE1999285}). We will now detail our main results on the impact of this phase transition on all the characteristics of the 
front describing the distribution of the rightmost position of this process.

\subsection{Intermittent phase}

 In  the intermittent phase, for $\delta+\gamma>1/2$, the rescaled variable $\zeta$ appearing in Eq. \eqref{late_t_cond} reads
\be
\zeta_{\rm i}=x-m(t)\;\;{\rm where}\;\;m(t)=v_{\rm i} t-\frac{3}{2\lambda_{\rm i}}\ln t+O(1)\;,\label{xi_int}
\ee
and the position of the front $m(t)$ is computed up to an unknown time-independent correction of $O(1)$ in this phase. Here and in the following, the subscripts $_{\rm i}$ and superscripts $^{\rm i}$ refer to the intermittent phase. The value of the velocity $v=v_{\rm i}$ of the TF appearing in Eq. \eqref{xi_int} is given by
\be
v_{\rm i}=v_0\sqrt{1-\left(\frac{r_s}{r_a}\right)^2}=\frac{2\sqrt{(1-\delta)(\delta+2\gamma)}}{2\gamma+1}v_0,\label{v_TW2} 
\ee
where $r_s=1/2-\delta-\gamma$ is the inverse life-time of clusters in the intermittent phase and $r_a=\gamma+1/2$ is the rate at which particles in the direction $\sigma$ produce particles in the opposite direction $-\sigma$, either by tumbling or branching. The parameter $\lambda_{\rm i}$ appearing in the correction to the front's position in Eq. \eqref{xi_int} characterises the tail of the TF solution
\be
F_{\sigma}(\zeta_{\rm i}\to +\infty)\approx a_{\sigma} \zeta_{\rm i} e^{-\lambda_{\rm i} \zeta_{\rm i}}\;,\label{as_int}
\ee
and reads
\be
\lambda_{\rm i}=\frac{r_a^2 v_{\rm i}}{r_s v_0^2}=\frac{1}{v_0}\frac{2\gamma+1}{2\gamma+2\delta-1}\sqrt{(1-\delta)(\delta+2\gamma)}\;.\label{l_star}
\ee
As the particles in this model have a finite maximal speed $v_0$, this tail is cut-off at the value $\zeta^{\rm c.o.}(t)$ for any finite $t$, such that $F_{\sigma}(\zeta_{\rm i}>\zeta^{\rm c.o.}(t))=0$, as seen in Fig. \ref{Fig_F_int}. The value of $\zeta^{\rm c.o.}(t)$ is simply obtained by inserting $x=v_0 t$ in Eq. \eqref{xi_int}. The TF solution approaches the value $1$ exponentially for $\zeta_{\rm i}\to -\infty$ \eqref{zeta_i_as} with the rate $\lambda$ in Eq. \eqref{rate_i}. 
Note that the main features of the TF in this phase are qualitatively similar as that of the TF which is solution of the Fisher-KPP equation. An important difference is that the TF at long time still depends explicitly on the direction $\sigma=\pm$ of the initial particle.

\subsection{Persistent phase}

In the persistent phase, corresponding to $\delta+\gamma<1/2$, the rescaled variable $\zeta$ appearing in Eq. \eqref{late_t_cond} reads
\be
\zeta_{\rm p}=x-v_0 t\;.\label{xi_per}
\ee
Here and in the following, the subscripts $_{\rm p}$ and superscripts $^{\rm p}$ refer to the persistent phase. The RTPs have in this model a finite maximal speed $+v_0$. This means that at any time, $x_{\max}(t)$ cannot be larger than $v_0 t$ and thus the TF has a finite edge $F_{\sigma}(\zeta_{\rm p}>0)=0$ (see Fig. \ref{Fig_F_per}). Moreover, there is a finite probability that $x_{\max}(t)=v_0 t$ if the initial direction is $\sigma=+$. For this event to occur, one should have $x_{\max}(\tau)=v_0 \tau$ at all times $\tau\in[0,t]$. It means that at all times up to $t$ there must to be a non-zero number of particles in the cluster emerging from the initial particle. On the other hand, if the initial particle starts in direction $\sigma=-$, it will take a finite amount of time for it to either tumble or create a right-moving offspring. This delay cannot be caught-up afterwards as the maximal velocity is $v_0$ and one has  $x_{\max}(t)<v_0 t$ at time $t$. This feature can be observed in Fig. \ref{Fig_F_per} where the cumulative distribution $Q_-(x,t)$ is continuous for $\zeta_{\rm p}=x-v_0 t=0$ while $Q_+(x,t)$ is discontinuous,
\begin{equation}
Q_+(v_0 t,t)=\Prob\left[x_{\max}(t)=v_0 t|+\right]\to1-2\gamma-2\delta\;.\label{fp_0}
\end{equation}
The definition of the front's position used in Eq. \eqref{xi_per} is $F_-(\zeta_{\rm p}=0)=0$. The velocity of the TF is $v=v_0$ and the correction to the front's position exactly $X(t)=0$.
In this phase, we obtain an exact expression for the TF solution $F^{\rm p}_{+}(\zeta_{\rm p})$, expressed in terms of its inverse function $Z_+^{\rm p}(f)$ in Eq. \eqref{inv_plus}. The travelling front with initial direction $\sigma=-$ can then be obtained from Eq. \eqref{Fp_Fm}. 
We also characterise analytically the asymptotic behaviours of the travelling front solution $F_{\sigma}^{\rm p}(\zeta_{\rm p})$ both for $\zeta_{\rm p}\to 0$ in Eqs. \eqref{Fm_sm} and \eqref{Fp_sm} and for $\zeta_{\rm p}\to -\infty$ in Eq. \eqref{as_per_infty} (see the rate in Eq. \eqref{decay_rate_per} and the pre-exponential constant in Eq. \eqref{const_per}). The finite edge and the discontinuity of the TF clearly arise from the persistence of the RTPs and are not observed for the BBM \cite{bramson1983convergence}. 

\subsection{At the transition}

At the transition, the travelling front has velocity $v_0$. The TF $F^{\rm t}_{\sigma}(\zeta_{\rm p})$, where the superscript $^{\rm t}$ refers to the transition, 
exhibits a finite edge for $\zeta_{\rm p}=0$. Starting in the state $\sigma=+$, the TF with initial direction $\sigma=+$ is discontinuous at $\zeta_{\rm p}=0$ for any finite time
, with
\be
Q_+(v_0 t,t)=\frac{2}{2+t}>Q_+(x>v_0 t,t)=0\;,
\ee
and becomes continuous only in the limit $t\to \infty$. In the special case where $\delta=0$ and $\gamma=1/2$, for which the full distribution $Q_{\sigma}(x,t)=F_{\sigma}^{\rm t}(\zeta_{\rm p})$ is described by the TF at sufficiently large time $t\gg 1$, we are able to obtain with good approximation the TF $F^{\rm t}_{\sigma}(\zeta_{\rm p})$ at large but finite time $t$ by computing its inverse function
\be
\frac{1}{v_0}Z_{+}^{\rm t}(f;t)=\frac{3}{2}\ln\left(\frac{2+t}{t}(1-f)\right)-\frac{f}{2}+\frac{1}{2+t}\;.\label{inv_t_res}
\ee
The TF with initial directions $\sigma=\pm$ satisfy with good approximation 
the algebraic relation $F_-^{\rm t}={F_+^{\rm t}}^2/(2-F_+^{\rm t})$ at these large times.
Finally, the function $Z_+^{\rm t}(f)$, inverse of the TF solution $F_+^{\rm t}(\zeta_{\rm p})$ can be obtained in the limit $t\to \infty$ for any $\gamma,\delta$ with $\gamma+\delta=1/2$. It is obtained by taking $t\to \infty$ in Eq. \eqref{inv_t_res} and reads $(r_a/v_0)Z_+^{\rm t}(f)=3/2\ln(1-f)-f/2$.
From this expression, we extract the asymptotic behaviours of the TF solution both for $\zeta_{\rm p}\to 0$ in Eqs. \eqref{Fp_t_sm} and \eqref{Fm_t_sm} and for $\zeta_{\rm p}\to -\infty$ in Eq. \eqref{F_t_inf}.

\section{Fokker-Planck equation and travelling front solution}\label{sec_eq}

\subsection{Fokker-Planck equation}

Our analysis starts with the derivation of a backward Fokker-Planck equation \cite{risken1996fokker} for the cumulative probability in Eq. \eqref{Q_max_def} (see \cite{ramola2015spatial} for a similar derivation for BBM). 
In this approach, an equation for $Q_{\s}(x,t+dt)$ is obtained by splitting the time interval $[0,t+dt]$ into $[0,dt]$ and $[dt,t+dt]$. One first considers all possible evolution of the initial particle during the first sub-interval $[0,dt]$. Then, given this initial evolution
, the equation is derived by considering all possible evolution in the interval $[dt,t+dt]$ which contribute to $x_{\max}(t+dt)\geq x$ . Initially, the maximum is $x_{\max}(t)=x_0=0$ and the corresponding cumulative probability is $Q_{\s}(x,0)=\Prob\left[x_{\max}(0)\geq x|\sigma\right]=\Theta(-x)$, where $\Theta(x)$ is the Heaviside step function. The evolution equation reads for $x>0$
\begin{align}
&Q_{\sigma}(x,t+dt)=\gamma dt Q_{-\sigma}(x,t)\label{evo_dt}
\\
&+\frac{dt}{2}\left[2Q_{\sigma}(x,t)-Q_{\sigma}(x,t)^2\right]\nn\\
&+\frac{dt}{2}\left[Q_{\sigma}(x,t)+Q_{-\sigma}(x,t)-Q_{\sigma}(x,t)Q_{-\sigma}(x,t)\right]\nn\\
&+\left[1-(1+\gamma+\delta)dt\right]Q_{\sigma}(x-\sigma v_0 dt,t)\nn\;. 
\end{align}
To understand this equation, consider the evolution of the initial particle during the first interval of time $dt$. It can either:\\
(i) Die with probability $\delta dt$, such that $x_{\max}(t)=0<x$ for any time $t\geq 0$, implying this process does not contribute to \eqref{evo_dt}.
\\
(ii) Tumble with probability $\gamma dt$. After time $dt$, the initial particle has direction $-\sigma$. Then $x_{\max}$ remains bounded to $[x,\infty)$ during $[dt,t+dt]$ with probability $Q_{-\sigma}(x,t)$, explaining the first line of Eq. \eqref{evo_dt}\\
(iii) Branch with probability $dt$. Its offspring is in the direction $\sigma_{\rm o}=\pm$ with probability $1/2$. There are now two independent branching RTP processes whose maximum $x_{\max}(t+dt)$ must remain bounded to $[x,\infty)$. This event happens if the maximum of either process remains bounded to $[x,\infty)$. The corresponding probability is $Q_{\sigma}(x,t)+Q_{\sigma_{\rm o}}(x,t)(1-Q_{\sigma}(x,t))$. The term for $\sigma_{\rm o}=\sigma$ corresponds to the second line of Eq. \eqref{evo_dt} and the term for $\sigma_{\rm o}=-\sigma$ to the third line\\
(iv) Move ballistically to position $\sigma v_0 dt$ with probability $1-(1+\gamma+\delta)dt$. The maximum $x_{\max}(t+dt)$ remains bounded to $[x,+\infty)$ with probability $Q_{\sigma}(x-\sigma v_0 dt,t)$, explaining the fourth line of Eq. \eqref{evo_dt}.\\
In the limit $dt\to 0$ and for $x> 0$, we obtain the two partial non-linear differential evolution equations for $\sigma=\pm$
\begin{align}
\partial_{t}Q_{\sigma}(x,t)=&-\sigma v_0\partial_{x} Q_{\sigma}(x,t)+r_a Q_{-\sigma}(x,t)-r_s Q_{\sigma}(x,t)\nn\\
&-\frac{Q_{\sigma}(x,t)}{2}\left[Q_{-\sigma}(x,t)+Q_{\sigma}(x,t)\right]\label{Q_max}\;,
\end{align}
where the rates $r_a,r_s$ read
\be
r_a=\gamma+\frac{1}{2}\;,\;\;r_s=\frac{1}{2}-\gamma-\delta\;.
\ee
As we will see in the following, the fact that $r_s>0$ in the intermittent phase, while $r_s<0$ in the persistent phase is crucial to understand the impact of the phase transition on the EVS of this process. We will now consider the behaviour of the cumulative probability in the large time limit and focus in particular on the travelling front.

\subsection{Travelling front solution}

After a typical time of order $O(1)$, the probability $P_0(t)=\Prob\left[N(t)=0\right]$ reaches its asymptotic value (See App. \ref{app_num} for details). Supposing the extinction of the process, and since 
there is a finite maximal velocity $v_0$, the process only has time to reach positions $x_{\max}(t)=O(1)$ before all RTP die. In the following, we focus our attention on the region $x\propto t\gg 1$, where the stationary distribution vanishes, i.e. 
$Q_{\sigma}^{\rm st}\left(x\right)\to 0$, while the travelling front solution $F_{\sigma}(\zeta)$ is non-zero. In the long time limit, using the asymptotic scaling form of 
the CDF in Eq. \eqref{late_t_cond} and assuming 
that at leading order the position of the front evolves ballistically $m(t)=v t +o(t)$, one can replace $\partial_t F_{\sigma}(\zeta) \to -v\partial_{\zeta}F_{\sigma}(\zeta)+o(1)$
and re-express the evolution of the CDF in Eq. \eqref{Q_max} in terms of the TF solution $F_\sigma(\zeta)$ as 
\begin{align}
&\left(v-\sigma v_0\right)\partial_{\zeta} F_{\sigma}(\zeta)=-r_a F_{-\sigma}(\zeta)+r_s F_{\sigma}(\zeta)\nn \\
&+\frac{r_a-r_s}{2} F_{\sigma}(\zeta)\left[F_{\sigma}(\zeta)+F_{-\sigma}(\zeta)\right]+o(1)\;.\label{TW_eq}
\end{align}
 This equation for the TF solution can be amenable to an alternative form, closer to the Fisher-KPP equation. To see this, we multiply by $(v+\sigma v_0)$ and take a derivative with respect to $\zeta$. It yields
\begin{widetext}
\begin{align}
&\left(v_0^2-v^2\right)\partial_{\zeta}^2 F_{\sigma}+2v r_s \partial_{\zeta} F_{\sigma}+(r_a^2-r_s^2)F_{\sigma}=\frac{r_a-r_s}{2}\left[(F_{\sigma}+F_{-\sigma})(r_s F_{\sigma}+r_a F_{-\sigma})-\left(v+\sigma v_0\right)\partial_{\zeta}\left(F_{\sigma}(F_{\sigma}+F_{-\sigma})\right)\right]\;,\label{Eq_KPP_style}
\end{align} 
\end{widetext}
where we have used the short-handed notation $F_{\sigma}\equiv F_{\sigma}(\zeta)$. This equation has two fixed-points, which are uniform solutions, $F_{\pm}=1$ which is stable and $F_{\pm}=0$ which is unstable. As time evolves, spatial areas that correspond to the unstable solution are invaded by the stable solution. This scenario is typical of non-linear equations such as the Fisher-KPP equation \cite{van2003front}. 
We now proceed to analyse the large time behaviour of the TF solution $F_{\sigma}\left(\zeta \right)$ in each of the two phases independently. 


\section{Intermittent phase}\label{sec_int_phase}

\begin{figure}
    \centering
    \includegraphics[width=0.475\textwidth]{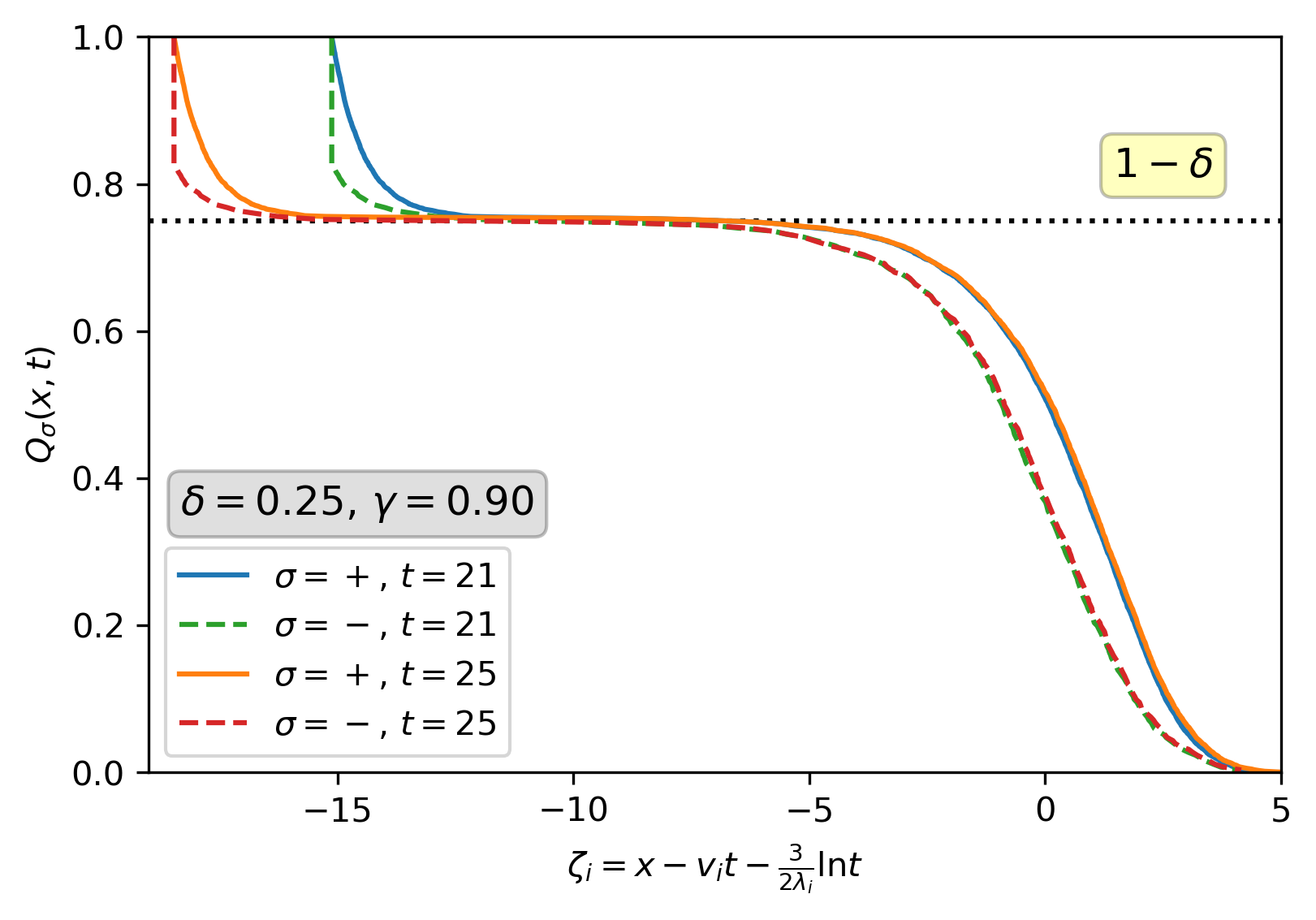}
    \caption{
    Plot of $Q_{\pm}(x,t)$ versus the rescaled position $\zeta_{\rm i}$ given in Eq. \eqref{xi_int} obtained from numerical simulations of the stochastic process in the intermittent phase ($\gamma+\delta>1/2$). The dashed horizontal line marks the value of $1-\delta$, separating the contribution of the stationary distribution $\delta\,Q_{\sigma}^{\rm st}\left(x\right)$ (above) from the TF contribution $(1-\delta)\,F_{\sigma}(\zeta_{\rm i})$ (below) as described in Eq. \eqref{late_t_cond}. The numerical results for the TF contributions with initial direction $\sigma=\pm$ and for $t=21,25$ exhibit a very good collapse on the same master curves. As particles have a finite maximal speed $+v_0$, the cumulative probability vanishes exactly $Q_{\pm}(v_0 t,t)=0$ at the rescaled position $\zeta^{\rm c.o.}(t)=(v_0-v_{\rm i}) t-3/(2\lambda_{\rm i})\ln t$.}
    \label{Fig_F_int}
\end{figure}


We first consider the behaviour of the TF solution in the intermittent phase. 

\subsection{Velocity of the travelling front}

For completeness, we first repeat the general arguments allowing to derive the velocity $v$ of the travelling front equation and show that in the intermittent phase $v_{\rm i}$ satisfies Eq. \eqref{v_TW2} \cite{HORSTHEMKE1999285, demaerel2019asymptotic, bouin2015propagation, bouin2014hyperbolic, fedotov1998traveling}. In the limit $\zeta\to \infty$, the function $F_{\sigma}(\zeta)$ approaches the fixed point $F_{\sigma}=0$ of Eq. \eqref{Eq_KPP_style}. Both its value and the value of 
its derivative are small in this regime ($-\partial_\zeta F_{\sigma}$ is a probability distribution, it is positive and goes to zero as $\zeta\to \infty$). We may neglect the non-linear terms in Eq. \eqref{Eq_KPP_style} and approximate the full solution $F_{\sigma}(\zeta)$ by the solution $F_{\sigma}^{\rm lin}(\zeta)$ of the linearised equation. In this linearised version of Eq. \eqref{Eq_KPP_style}, the right-hand-side of the equation vanishes while the left-hand-side remains identical, yielding 
\be
\left(v_0^2-v^2\right)\partial_{\zeta}^2 F_{\sigma}^{\rm lin}+2r_s v \partial_{\zeta}F_{\sigma}^{\rm lin}+(r_a^2-r_s^2)F_{\sigma}^{\rm lin}=0\;.\label{F_lin}
\ee
The full non-linear solution, which shares the same asymptotic behaviour as the linear solution $F_{\sigma}(\zeta\to \infty)\sim F_{\sigma}^{\rm lin}(\zeta\to \infty)$ must be a decreasing function of $\zeta$ as it is a CDF. This means that we need to consider only the case where 
\be
v\geq v_{\rm i}=v_0\sqrt{1-\left(\frac{r_s}{r_a}\right)^2}\;.\label{v_i_2}
\ee
Note that the finite maximum velocity $v_0$ of the RTPs in our model imposes the other condition $v\leq v_0$.
In the opposite scenario $v<v_{\rm i}$, the linear solution is oscillating as $\zeta\to \infty$. The solution of the linear equation in Eq. \eqref{F_lin} reads
\be
F_{\sigma}^{\rm lin}(\zeta)=\begin{cases}
\displaystyle A_{\sigma}e^{-\lambda_+ \zeta}+B_{\sigma}e^{-\lambda_- \zeta} \;,\;\;v\geq v_{\rm i}\;,\\
\\
\displaystyle (a_{\sigma}\zeta+b_{\sigma})e^{-\lambda_{\rm i} \zeta}\;,\;\;v=v_{\rm i}\;,
\end{cases}\label{lin_sol}
\ee
where the decay rates $\lambda_{\pm}$ and $\lambda_{\rm i}$ are given by
\be
\lambda_{\pm}=\frac{r_s v}{v_0^2-v^2}\left[1\pm \frac{v_0}{v}\sqrt{ \frac{v^2-v_{\rm i}^2}{v_0^2-v_{\rm i}^2}}\right]\,;\,\lambda_{\rm i}=\frac{r_a^2 v_{\rm i}}{r_s v_0^2}\,,\label{l_pm}
\ee
with, in particular, $\lambda_{\rm i}=\lambda_{\pm}$ for $v=v_{\rm i}$. Let us first mention that for $r_s<0$, which is the case in the persistent phase, all the rates are negative $\lambda_{\pm},\lambda_{\rm i}<0$. The TF solution $F_{\sigma}(\zeta)$ would be exponentially increasing (instead of decaying) as $\zeta\to \infty$, which is un-physical: $F_{\sigma}(\zeta)$ is a cumulative probability and is therefore bounded $0\leq F_{\sigma}(\zeta)\leq 1$. The method used here therefore only applies to the intermittent phase where $r_s>0$. Since $\lambda_+\geq \lambda_-$, the solution of the linear equation will decay at infinity as $F_{\sigma}^{\rm lin}(\zeta\to \infty)\propto e^{-\lambda_{-}\zeta}$. Considering only the linear behaviour, any pair $(v,\lambda_-)$ represents a physical solution. The correct and unique solution to the non-linear equation is obtained by 
recalling that $F_{\sigma}(\zeta)$ is the TF solution at long time describing the CDF $Q_{\sigma}(x,t)$. The initial condition for this distribution is a step function $Q_{\sigma}(x,0)=\Theta(-x)$ and thus decays faster than exponentially as $x\to \infty$. The tail $F_{\sigma}(\zeta\to \infty)\propto e^{-\lambda_{-}\zeta}$ of the solution, even at very large time, keeps trace of this initial condition such that the exponential decay rate $\lambda_-$ must be the largest possible \cite{van2003front,majumdar2003extreme}. 
Here, the largest real value of $\lambda_-$ is achieved for $v=v_{\rm i}$ and given by $\lambda_{\rm i}$. Inserting in Eq. \eqref{l_pm} the values of $r_s,r_a$, we recover the expression of $\lambda_{\rm i}$ in Eq. \eqref{l_star}. 

The expression for the velocity in the intermittent phase in Eq. \eqref{v_i_2} has a clear interpretation in terms of the rates $r_s$ and $r_a$. It increases with the rate $r_a$ at which left-moving RTPs (that cannot contribute to the growth of $x_{\max}(t)$) tumble or branch into right-moving RTPs. On the other hand, it decreases with the rate $r_s$, which is the inverse life-time of a cluster of particles.
Let us now turn to the correction $X(t)$ to the front's position defined in Eq. \eqref{front_pos}. This subject has been extensively studied for the Fisher-KPP equation and related non-linear equations and arises in a number of physical problems \cite{ebert2000front,berestycki2017exact,berestycki2018new}. 

\subsection{Correction to the front's position}

We have seen in the previous section that, at leading order, the front's position propagates ballistically $m(t)=v_i t+o(t)$. We now consider the behaviour of the correction $X(t)=(x-v_{\rm i}t)-\zeta_{\rm i}$ in the intermittent phase as $t\to \infty$. To derive this behaviour, we will again compare the behaviour of the full non-linear solution $F_{\sigma}(\zeta_{\rm i})$ as $\zeta_{\rm i}\to \infty$ to the behaviour of a linearised solution. We will consider here the solution $q_{\sigma}^{\rm lin}(x,t)$ of Eq. \eqref{Q_max}, linearised close to the fixed point $Q_{\sigma}=0$. It satisfies the telegraphic equation 
\be
\partial_t^2 q_{\sigma}^{\rm lin}+2r_s\partial_t q_{\sigma}^{\rm lin}=v_{0}^2\partial_x^2 q_{\sigma}^{\rm lin}+(r_a^2-r_s^2) q_{\sigma}^{\rm lin}\;.\label{lin_eq_x_t}
\ee

First, let us check that the leading asymptotic behaviour of the TF is indeed given by $F_{\sigma}^{\rm i}(\zeta_{\rm i}\to \infty)\approx a_{\sigma}\zeta_{\rm i} e^{-\lambda_{\rm i}\zeta_{\rm i}}$ and thus that $a_{\sigma}>0$. To show this, let us suppose that this asymptotic behaviour is indeed correct and compute the value of $a_{\sigma}$. Multiplying Eq. \eqref{Eq_KPP_style} by $(v_0^2-v_{\rm i}^2)^{-1} e^{\lambda_{\rm i}\zeta}$ and integrating with respect to $\zeta$ over the real line, one can show that 
the left-hand side of the equation simply reads 
\be
\int_{-\infty}^{\infty}\partial_{\zeta_{\rm i}}^2(e^{\lambda_{\rm i}\zeta_{\rm i}}F_{\sigma}^{\rm i}(\zeta_{\rm i}))d\zeta_{\rm i}=\left. \partial_{\zeta_{\rm i}}(e^{\lambda_{\rm i}\zeta_{\rm i}}F_{\sigma}^{\rm i}(\zeta_{\rm i}))\right|_{\zeta_{\rm i}\to +\infty}=a_{\sigma}\;,
\ee
where we used the limit $F_{\sigma}^{\rm i}(\zeta_{\rm i}\to -\infty)=1$ ( $F_{\sigma}^{\rm i}$ is a complementary cumulative distribution) and our hypothesis for the asymptotic behaviour of $F_{\sigma}^{\rm i}(\zeta_{\rm i}\to \infty)$. The right-hand side can be re-expressed after using the integration by parts $\int e^{\lambda_{\rm i}\zeta}\partial_\zeta (F_{\sigma}(F_{\sigma}+F_{-\sigma}))d\zeta=-\lambda_{\rm i} \int e^{\lambda_{\rm i}\zeta}F_{\sigma}(F_{\sigma}+F_{-\sigma})d\zeta$. It yields an exact expression for $a_{\sigma}$
\begin{align}
a_{\sigma}=\int_{-\infty}^{\infty}d\zeta\,&\frac{r_a(r_a-r_s)(r_a^2-r_s^2) e^{\lambda_{\rm i}\zeta}}{2r_s^3} [F_{\sigma}^{\rm i}+F_{-\sigma}^{\rm i}]\label{a_ugly}\\
&\times\left[(r_a+\sigma \sqrt{r_a^2-r_s^2})F_{\sigma}+r_s F_{-\sigma}\right]\;.\nn
\end{align}
This expression is strictly positive in the intermittent phase, as the rates satisfy $r_a>r_s>0$ and $0\leq F_{\sigma}\leq 1$. Our hypothesis for the leading behaviour of the non-linear solution $F_{\sigma}^{\rm i}(\zeta_{\rm i}\to \infty)\approx a_{\sigma}\zeta_{\rm i} e^{-\lambda_{\rm i}\zeta_{\rm i}}$ is therefore valid. 

The behaviour of the leading correction $X(t)$ to the front's position for a non-linear equation built on the telegraphic equation has been investigated in \cite{gallay2000stability,ebert2000front}. If the TF solution behaves asymptotically as $F_{\sigma}^{\rm i}(\zeta_{\rm i}\to \infty)\approx a_{\sigma}\zeta_{\rm i} e^{-\lambda_{\rm i}\zeta_{\rm i}}$, the leading correction takes the scaling form
\be
X(t)\approx \frac{3}{2\lambda_{\rm i}}\ln t+O(1)\;,\label{corec_X}
\ee
where $\lambda_{\rm i}$ depends explicitly on the details of the equation but the $(3/2)\ln t$ is universal and is also obtained e.g. in the case of BBM.
We show here that this result is indeed correct for our particular case. In the asymptotic regime $\zeta_{\rm i}\to \infty$, we expect that the functions $F_{\sigma}^{\rm i}(\zeta_{\rm i})\sim q_{\sigma}(x,t)$ share the same asymptotic behaviour. We therefore need to obtain the asymptotic behaviour of $q_{\sigma}(x,t)$ for $t\gg 1$ and $x- v_{\rm i} t\gg 1$. We first remove the exponential tail by considering $\tilde q_{\sigma}(x,t)=e^{\lambda_{\rm i}(x-v_{\rm i}t)} q_{\sigma}(x,t)$, which satisfies the equation
\be
\partial_t^2 \tilde q_{\sigma}^{\rm lin}+2(r_s+\lambda_{\rm i} v_{\rm i})\partial_t \tilde q_{\sigma}^{\rm lin}=v_{0}^2\partial_x^2 \tilde q_{\sigma}^{\rm lin}-2\lambda_{\rm i}v_0^2\partial_x \tilde  q_{\sigma}^{\rm lin}\;.\label{q_tilde}
\ee
Note that the factor $x-v_{\rm i}t=\zeta_{\rm i}+X(t)$ can conveniently be expressed in terms of the correction $X(t)$ to the front's position.
In the large $t$ limit, one can show for this telegraphic equation that the solution is a scaling function of $u=(x-v_{\rm i}t)/\sqrt{t}=O(1)$. The scaling form leading to the correct asymptotic behaviour reads 
\be
\tilde q_{\sigma}^{\rm lin}(x,t)=\frac{1}{t}g\left(\frac{x-v_{\rm i}}{\sqrt{t}}\right)\;.
\ee
Inserting this scaling form in Eq. \eqref{q_tilde} and in the regime $t\gg 1$ with $u=0(1)$, the scaling function $g(u)$ satisfies
\be
D_{\rm eff} g''(u)+u g'(u)+2g(u)=0\;,\;\;D_{\rm eff}=\frac{v_0^2 r_a^3}{r_a^4}\;.
\ee
The solution which leads to a correct matching is given by
\be
g(u)= A\, u\, e^{-\frac{u^2}{2D_{\rm eff}}}\;.
\ee
Inserting this scaling function into the solution of the linear equation $q^{\rm lin}_{\sigma}(x,t)$, one obtains at large time and in the regime $\sqrt{t}\gg x-v_{\rm i}t=\zeta_{\rm i}+X(t)\gg \ln t$,
\be
q_{\sigma}^{\rm lin}(x,t)\approx A \frac{(\zeta_{\rm i}+X(t))}{t^{3/2}}e^{-\lambda_{\rm i}(\zeta_{\rm i}+X(t))}\;.
\ee
Comparing with the asymptotic behaviour $F_{\sigma}^{\rm i}(\zeta_{\rm i}\to \infty)\approx a_{\sigma}\zeta_{\rm i} e^{-\lambda_{\rm i}\zeta_{\rm i}}$ and supposing $\zeta_{\rm i}\gg X(t)$, one recovers that at leading order, the correction to the front's position $X(t)$ is indeed given by Eq. \eqref{corec_X}. The computation of the corrections beyond this log correction is in general a complicated task \cite{berestycki2017exact,berestycki2018new}.


\subsection{Asymptotic behaviour for $\zeta_{\rm i}\to -\infty$}

We already characterised in Eq. \eqref{as_int} the asymptotic behaviour of the TF solution for $\zeta_{\rm i}\to +\infty$ and used this result in the previous section to obtain the correction to the front's position. 
We will now consider the asymptotic behaviour in the opposite limit $\zeta_{\rm i}\to -\infty$. Since $F_{\sigma}^{\rm i}(\zeta_{\rm i})$ is a (complementary) cumulative distribution, it converges to the stable fixed point $F_{\sigma}^{\rm i}(\zeta_{\rm i})\to 1$ at leading order as $\zeta_{\rm i}\to -\infty$. We will now consider the first order correction to this behaviour by linearising the equation for the TF solution \eqref{TW_eq} close to the stable fixed point, introducing  $f_{\sigma}(\zeta_{\rm i})=1-F_{\sigma}(\zeta_{\rm i})\ll 1$ as $\zeta_{\rm i}\to -\infty$. It yields
\begin{align}
&2\left(v_{\rm i}-\sigma v_0\right)\partial_{\zeta_{\rm i}} f_{\sigma}=(3r_a-r_s) f_{\sigma}-(r_a+r_s) f_{-\sigma}\;,
\end{align}
where $f_{\sigma}\equiv f_{\sigma}(\zeta_{\rm i})$.
Note that we used that the derivative $\partial_{\zeta_{\rm i}} f_{\sigma}$ is the probability distribution of $x_{\max}(t)$ conditioned on the survival of the process and approaches zero as $\zeta_{\rm i}\to -\infty$. Solving the linear differential equation with the correct asymptotic behaviour $f_{\sigma}(\zeta_{\rm i}\to -\infty)\to 0$, one obtains
\be
F_{\sigma}(\zeta_{\rm i}\to -\infty)=1-b_{\sigma}e^{\lambda \zeta_{\rm i}}\;,\label{zeta_i_as}
\ee
with the decay rate
\begin{align}
\lambda=&\frac{r_a}{2r_s^2 v_0}\sqrt{(r_a-r_s)(3r_a+r_s)(3r_a^2+r_s^2)}\label{rate_i}\\
&-(3r_a-r_s)\frac{r_a\sqrt{r_a^2-r_s^2}}{2r_s^2 v_0}\;.\nn
\end{align}
Note that a similar exponential decay of the solution observed for the solution of the Fisher-KPP equation close to the stable fixed point. 

\subsection{Brownian limit}

We have seen in the previous sections that the TF solution in the intermittent phase shares a number of features with the TF describing the maximum of a BBM, solution of the Fisher-KPP equation. These qualitative features include the asymptotic behaviours of the solution as $\zeta_{\rm i}\to \infty$ in Eq. \eqref{as_int} and as $\zeta_{\rm i}\to -\infty$ in Eq.  \eqref{zeta_i_as} together with the log-correction to the front's position as seen in Eqs. \eqref{xi_int} and \eqref{corec_X}. We are now going to show that there is a limit in which this analogy is also quantitative and the TF in the intermittent phase converges exactly to the solution of the Fisher-KPP equation.
It has been well-established that in the limit $\gamma\to \infty$, $v_0\to \infty$ with $D=\frac{v_0^2}{2\gamma}=O(1)$, the RTP converges to a Brownian motion. In this limit, we first use that both $r_a\approx r_s\approx \gamma$ at leading order with $r_a-r_s=1-\delta$ to rewrite the velocity of the TF in Eq. \eqref{v_TW2} and the exponential decay rates in Eq. \eqref{l_star} and Eq. \eqref{rate_i} as
\begin{align}
&v_{\rm i}\to v_{\rm BBM}=2\sqrt{D(1-\delta)}\;,\\
&\lambda_{\rm i}\to \frac{v_{\rm BBM}}{2D}\;,\;\;\lambda\to (\sqrt{2}-1)\frac{v_{\rm BBM}}{2D}\;.
\end{align}
We recover exactly in this limit the values of velocity and decay rates for the BBM. This can be understood further by taking the same limit in the TF equation \eqref{Eq_KPP_style}, yielding
\be
\frac{1}{\lambda_{\rm i}^{2}} F_{\sigma}''+\frac{2}{\lambda_{\rm i}} F_{\sigma}'+F_{\sigma}-\frac{(F_{\sigma}+F_{-\sigma})^2}{4}=0\;.\label{Brown_lim_1}
\ee
Note that the non-linear term is symmetric in $\sigma$. Therefore the function $f(\zeta_{\rm i})=F_{+}(\zeta_{\rm i})-F_{-}(\zeta_{\rm i})$ is solution of a second order linear differential equation. Since the functions $F_{\pm}(\zeta_{\rm i})$ both converge to the same fixed point for $\zeta_{\rm i}\to -\infty$ and $\zeta_{\rm i}\to +\infty$, the function $f(\zeta_{\rm i}\to \pm \infty)=0$ is zero in both limits. It yields that $f(\zeta_{\rm i})=0$ for all $\zeta_{\rm i}$. One thus has $F_{+}(\zeta_{\rm i})=F_{-}(\zeta_{\rm i})\equiv F(\zeta_{\rm i})$ for all $\zeta_{\rm i}$. This is not surprising as in this limit, particles tumble from one state to the other instantaneously, such that the direction of the initial particle becomes irrelevant. Rescaling the position $\zeta_{\rm i}\to \xi =\lambda_{\rm i}\zeta_{\rm i}$, Eq. \eqref{Brown_lim_1} is amenable to the Fisher-KPP equation
\be
F''(\xi)+2F'(\xi)+F(\xi)-F^2(\xi)=0\;.\label{Brown_lim_FKPP}
\ee
The TF solution for the branching RTP process converges in this limit to the TF solution for BBM as one should expect. Note that taking the limit $\gamma\to \infty$, $v_0\to \infty$ with $D=\frac{v_0^2}{2\gamma}=O(1)$ directly in the initial equation for the CDF \eqref{Q_max}, one can show that the full CDF $Q_{\sigma}(x,t)\to Q_{\rm BBM}(x,t)$ converges to that of a Brownian motion \cite{ramola2015branching}.
This convergence of the solution for RTP to the solution for BBM thus also applies to the stationary distribution $Q_{\sigma}^{\rm st}(x)$ obtained in case of extinction.  

\section{Persistent phase}\label{sec_per_phase}

We now consider the persistent phase where $r_s=\gamma+\delta-1/2<0$. In this phase,  we will see that the velocity of the travelling front is maximal and equal to $v_0$ and that the TF has a finite edge beyond which it vanishes exactly. Before considering the general case, we first show that this behaviour indeed exists in the persistent phase by considering the special case of infinite persistence where there is no tumbling $\gamma=0$ and no death $\delta=0$ and particles have simple ballistic motion. 


\subsection{Infinite persistence ($\gamma=\delta=0$)}

\begin{figure}
    \centering
    \includegraphics[width=0.475\textwidth]{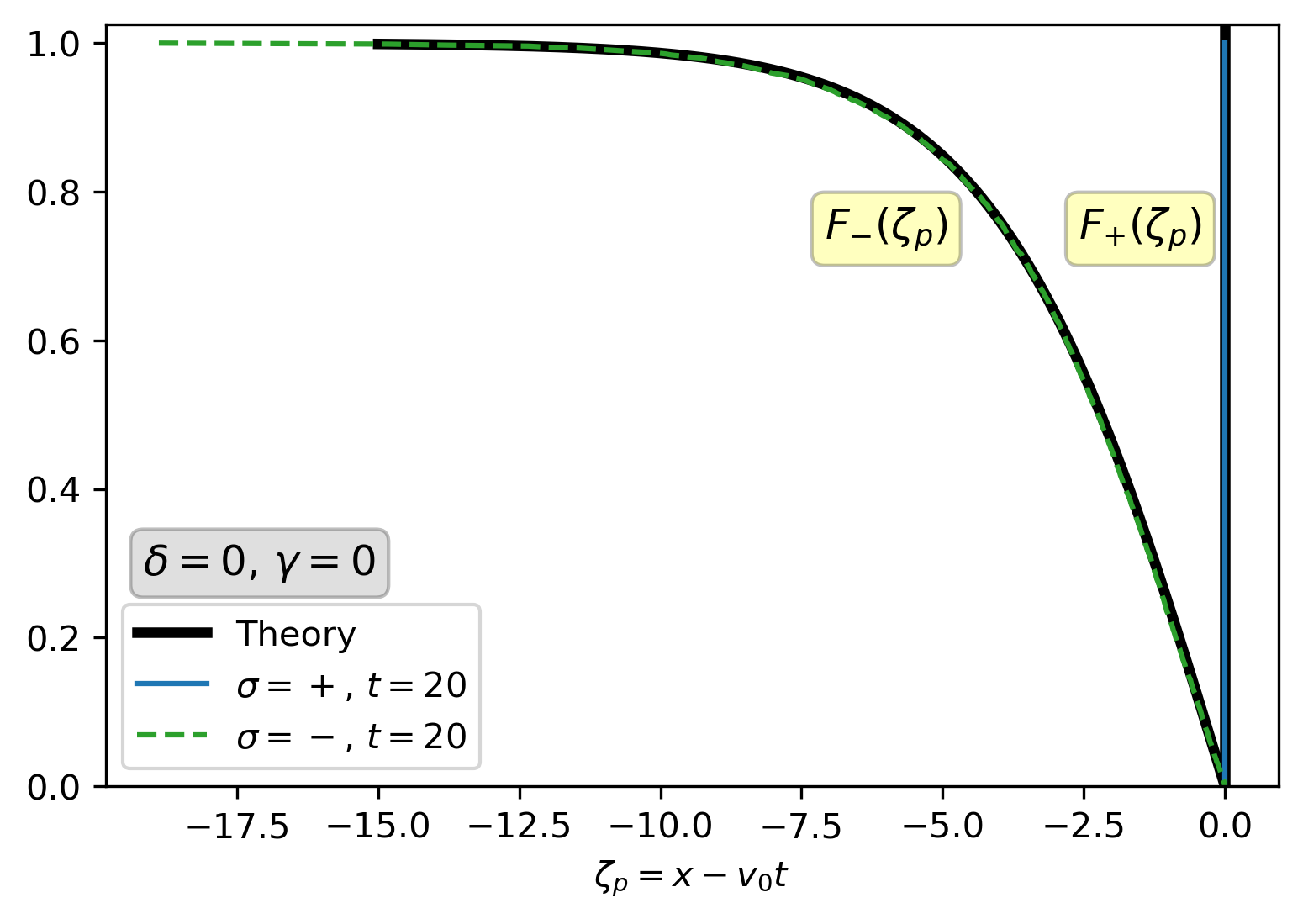}
    \caption{Plot of $Q_{\pm}(x,t)$ versus the rescaled position $\zeta_{\rm p}=x-v_0 t$ obtained from numerical simulation of the stochastic process for infinite persistence ($\gamma=\delta=0$). There is no stationary state contribution to $Q_{\pm}(x,t)$. The numerical results for the TF solution with initial direction $\sigma=\pm$ and for $t=20$ present a perfect collapse on the analytical results $F_{+}^{\rm p}(\zeta_{\rm p})=\Theta(-\zeta_{\rm p})$ and $F_{-}^{\rm t}(\zeta_{\rm p})$ given in Eq. \eqref{triv_fm}.
    }
    \label{Fig_F_inf_per}
\end{figure}

We first consider the simple case where there is no death $\delta=0$ and no tumbling $\gamma=0$. As there is no death, there is no steady state solution in Eq. \eqref{late_t_cond} and the full distribution is given at long time by the TF solution $Q_\sigma(x,t)=F_{\sigma}(\zeta_{\rm p})$, where $\zeta_{\rm p}=x-v_0 t$. Particles have ballistic motion with speed $\pm v_0$ for all time. It is then simple to realise that the maximum $x_{\max}(t)$ of the process is simply given by the position of the first right-moving particle to appear in the process. Starting in direction $\sigma=+$, the maximum of the process is the position of the initial particle and it yields that $x_{\max}(t)=v_0 t$ for all times $t$. The corresponding distribution retains its  
initial step profile at all times
\be
Q_{+}(x,t)=F_+^{\rm p}(\zeta_{\rm p}=x-v_0 t)=\Theta(-\zeta_{\rm p}).\label{triv_fp}
\ee
For $\delta,\gamma=0$, one has $r_a=-r_s=1/2$ and the equation for the CDF reads
\be
(\partial_{t}+\sigma v_0 \partial_{x})Q_{\sigma}=\frac{(1-Q_{\sigma})}{2}\left[Q_{-\sigma}+Q_{\sigma}\right]\;.\label{Q_inf_per}
\ee
It is then trivial to check that $Q_{+}(x,t)=\Theta(v_0 t-x)$ is a solution of Eq. \eqref{Q_inf_per}. 
Note that taking explicitly the boundary condition $Q_{\sigma}(x<0,t)=1$ into account, the distribution reads
\be
Q_{\sigma}(x,t)=\Theta(-x)+\Theta(x)F_{\sigma}(\zeta_{\rm p}=x-v_0 t)\;.\label{CDF_inf_per_m}
\ee
This expression is valid throughout the persistent phase in the particular case where $\delta=0$.
The TF solution $F_{-}(\zeta_{\rm p})$ can be computed exactly by inserting this form in Eq. \eqref{Q_inf_per} and solving the non-linear differential equation. However a more physical picture is obtained using the fact that $x_{\max}(t)=v_0(t-2\tau_{+})$, where $\tau_{+}$ is the stochastic time at which a right-moving particle first appears in the process.
The probability that at time $\tau$, there are $n$ left-moving and no right-moving particle is the probability that $\tau_+\geq\tau$ conditioned on the number of particles to be $n$. This probability is simply obtained as the probability that there are $n$ particles in the process (See App. \ref{app_num} for details) times the probability $2^{1-n}$ 
that all the $n-1$ offspring are left-moving
\be
\Prob\left[\tau_+\geq \tau|N(t)=n\right]=e^{-\tau}\left(\frac{1-e^{-\tau}}{2}\right)^{n-1}\;.
\ee
The cumulative probability of $\tau_+$ is then obtained by summing over all possible values of $n\geq 1$. Noting finally that $\Prob\left[x_{\max}(t)\geq v_0 t+\zeta_{\rm p}\right]=1-\Prob\left[\tau_+\geq -\zeta_{\rm p}/2\right]$ as $x_{\max}(t)=v_0(t-2\tau_{+})$, the TF solution reads
\begin{align}
F_{-}(\zeta_{\rm p})&=1-\sum_{n=1}^{\infty}\Prob\left[\tau_+\geq -\frac{\zeta_{\rm p}}{2} | N(t)=n\right]\nn\\
&=-\tanh\left(\frac{\zeta_{\rm p}}{4v_0}\right)\;.\label{triv_fm}
\end{align}
One can check that after inserting this TF solution in the full CDF in Eq. \eqref{CDF_inf_per_m}, $Q_{-}(x,t)$ satisfies Eq. \eqref{Q_inf_per}.
In Fig. \ref{Fig_F_inf_per}, we show a comparison between the TF obtained via numerical simulation of the process and our analytical results, showing excellent agreement. From Eq. \eqref{triv_fm}, we can extract the asymptotic behaviour for $\zeta_{\rm p}\to 0$
\be
F_{-}(\zeta_{\rm p}\to 0)=-\frac{\zeta_{\rm p}}{4v_0}+O(\zeta_{\rm p}^3)\;,\label{as_0_inf_per}
\ee
and for $\zeta_{\rm p}\to -\infty$,
\be
F_{-}(\zeta_{\rm p}\to -\infty)\approx 1-2e^{\zeta_{\rm p}/(2v_0)}\;.\label{as_inf_inf_per}
\ee
Finally, the moments of the maximum can be computed exactly. In the large $t$ limit, the rescaled random variable $\zeta_{\max}=x_{\max}(t)-v_0 t$ is independent of time and its cumulative distribution is precisely the TF solution $F_{\sigma}^{\rm p}(\zeta_{\rm p})$. The moments of $\zeta_{\max}$ read
\be
\moy{\zeta_{\max}^n}_{-}=\int_{-\infty}^{0}\frac{2n\zeta^{n-1}d\zeta}{1+e^{-\frac{\zeta}{2v_0}}}=(-v_0)^n 2 (2^n-2)n!{\cal Z}(n)\;,\label{mom_triv}
\ee
where ${\cal Z}(n)$ is the Riemann Zeta function, starting in direction $\sigma=-$ and $\moy{\zeta_{\max}^n}_{+}=0$ starting in direction $\sigma=+$.

We have seen in this section that in this simple case,
where there is no tumbling nor death, the TF exhibits 
a finite edge for $\zeta_{\rm p}=x-v_0 t=0$, beyond which it vanishes. In direction $\sigma=+$, the TF is discontinuous for $\zeta_{\rm p}=0$ as there is a finite probability (equal to $1$ here) that the cluster whose parent is the initial particle contains a non-zero number of particles at time $t$. The 
position of the edge can be computed exactly in this simple case.
We are going to show in the following 
that these features extend for any $\gamma$ and $\delta$ throughout the persistent phase ($\gamma+\delta<1/2$). We will in particular be able to obtain an exact analytical expression for the TF and characterise both its asymptotic behaviours and the moments of the distribution as we have obtained here in this simpler situation.

\subsection{Discontinuity of the solution for $x=v_0 t$}

First, we show that the discontinuity of the solution for $\zeta_{\rm p}=0$, i.e. $x=v_0 t$, is generic in the persistent phase. Replacing  $v\to v_0$ in the evolution of the TF Eq. \eqref{TW_eq} specified to $\sigma=+$, we derive an exact algebraic relation between the TF solutions $F_{+}^{\rm p}(\zeta_{\rm p})$ and $F_{-}^{\rm p}(\zeta_{\rm p})$,
\be
\kappa F_{+}^{\rm p}+ F_{-}^{\rm p}=\frac{1+\kappa}{2}F_{+}^{\rm p}\left(F_{+}^{\rm p}+F_{-}^{\rm p}\right)\;,\label{f_p_al}
\ee
where we have defined
\be
0<\kappa=-\frac{r_s}{r_a}=\frac{1-2\gamma-2\delta}{1+2\gamma}<1\;.\label{kappa}
\ee
Using this equation, one can express $F_{+}^{\rm p}(\zeta_{\rm p})$ as a function of $F_{-}^{\rm p}(\zeta_{\rm p})$ and $\kappa$ as
\be
F_{+}^{\rm p}=\frac{\kappa}{1+\kappa}-\frac{F_{-}^{\rm p}}{2}+\sqrt{\left(\frac{F_{-}^{\rm p}}{4}+\frac{(2-\kappa)}{2(1+\kappa)}\right)^2- \frac{4(1-\kappa)}{(1+\kappa)^2}}\;.\label{Fp_Fm}
\ee
Note that the previous section corresponds to the case where $\kappa=1$ with $r_a=1/2$ and in this limit $F_{+}^{\rm p}(\zeta_{\rm p})=1$ and does not depend on $F_{-}^{\rm p}(\zeta_{\rm p})$.
We will now use this equation to show that the function $F_{+}^{\rm p}(\zeta_{\rm p})$ is discontinuous for $\zeta_{\rm p}=0$ throughout the persistent phase.
Let us first remind that in our model, the RTP have a finite maximum velocity $+v_0$. It is therefore impossible for any RTP to be at a position larger than $v_0 t$ at time $t$. This yields trivially that
\be
F_{\sigma}^{\rm p}(\zeta_{\rm p}=x-v_0 t>0)\propto \Prob\left[x_{\max}(t)>v_0 t\right]=0\;.\label{prob_more}
\ee
We now consider the value of $F_{\sigma}(\zeta_{\rm p}=0)$. As we argued earlier, starting in the direction $\sigma=-$, it takes a finite amount of time for the initial left-moving particle to either tumble or create a right-moving offspring. One must therefore have $x_{\max}(t)< v_0 t$, even in the persistent phase. As $\zeta_{\rm p}=x-v_0 t$, this yields that 
$F_{-}^{\rm p}(\zeta_{\rm p}=0)=0$, as already seen for $\delta=\gamma=0$ in Eq. \eqref{triv_fm}.
Using this result by setting $F_{-}^{\rm p}=0$ in Eq. \eqref{Fp_Fm}, we obtain instead that
\be
F_{+}^{\rm p}(\zeta_{\rm p}=0)=\frac{2 \kappa}{1+\kappa}=\frac{1-2\gamma-2\delta}{1-\delta}\;.
\ee 
Inserting this result in the expression of the full CDF in Eq. \eqref{late_t_cond}, we obtain
\begin{align}
Q_{+}(v_0 t,t)&=\Prob\left[x_{\max}(t)=v_0 t\right]\label{Q_v0}\\
&=(1-\delta)F_{\sigma}^{\rm p}(0)=1-2\gamma-2\delta\;.\nn
\end{align}
Note that we used Eq. \eqref{prob_more} to rewrite $\Prob\left[x_{\max}(t)\geq v_0 t\right]=\Prob\left[x_{\max}(t)=v_0 t\right]$. As seen in Fig. \ref{Fig_F_per}, this discontinuity is confirmed by numerical simulation of the process.
This result can be interpreted as the probability that the cluster emerging from the initial particle contains a non-zero number of RTPs at time $t$. The position of this cluster of particles at any time $t$ is simply given by $v_0 t$, hence $x_{\max}(t)=v_0 t$. At $t=0$, there is only one particle, the initial one, in this cluster. The rate of branching within the cluster is $\frac{1}{2}$ while the rate at which particles leave the cluster, via tumbling or death, 
is $\delta+\gamma$. 
In the long time limit, the probability that there is a positive number of particles in this cluster converges to $1-2\delta -2\gamma$ (See App. \ref{app_num} for details), resulting in 
Eq. \eqref{Q_v0}. Now that we have obtained an exact expression for the discontinuity of the solution, we are going to show that there is an exact analytical expression for the function $Z_{+}^{\rm p}={F_{+}^{\rm p}}^{-1}$, i.e. the inverse function of the TF solution $F_{+}^{\rm p}(\zeta_{\rm p})$.
 
\subsection{Exact inverse functions}

\begin{figure}
    \centering
    \includegraphics[width=0.475\textwidth]{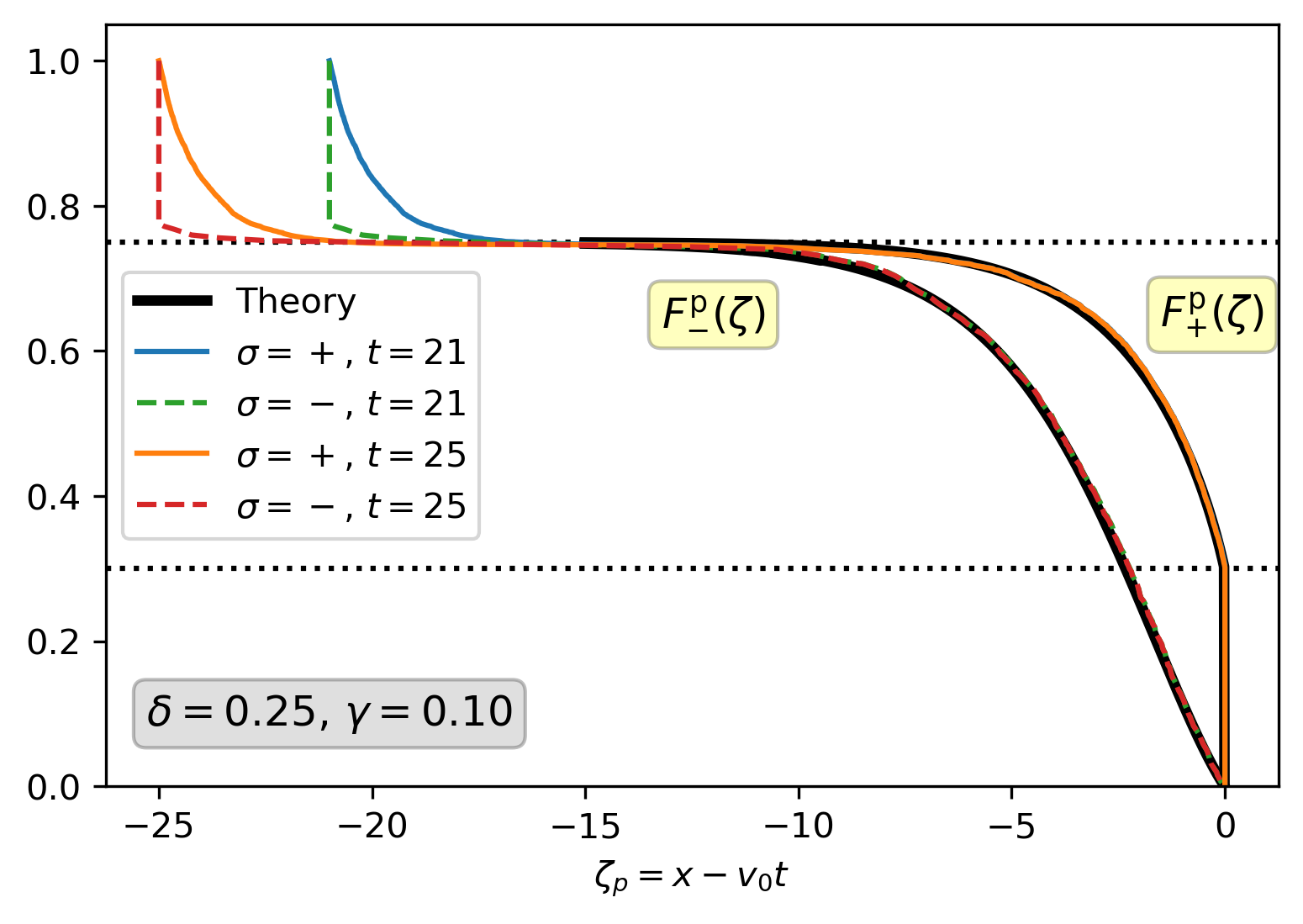}
    \caption{
    Plot of $Q_{\pm}(x,t)$ versus the rescaled position $\zeta_{\rm p}=x-v_0 t$ obtained from numerical simulation of the stochastic process in the persistent phase ($\gamma+\delta<1/2$). The upper dashed horizontal line marks the value of $1-\delta$, separating the contribution from the stationary state $\delta\,Q_{\sigma}^{\rm st}\left(x\right)$ (above) from the contribution of the TF  $(1-\delta)\,F_{\sigma}(\zeta_{\rm i})$ (below).  The numerical results for the TF contributions with initial direction $\sigma=\pm$ and for $t=21,25$ present a perfect collapse on the analytical results $F_{\sigma}^{\rm p}(\zeta_{\rm p})$ obtained by functional inversion of Eq. \eqref{inv_plus} and from Eq. \eqref{Fp_Fm}. Note the discontinuity for $\zeta_{\rm p}=0$ such that $Q_{+}(v_0 t+0_-,t)=1-2\delta-2\gamma>Q_{+}(v_0 t+0_+,t)=0$ marked by the lower horizontal line.}
    \label{Fig_F_per}
\end{figure}

To obtain the expression of the inverse function of $F_{\sigma}^{\rm p}(\zeta_{\rm p})$, we start by considering the differential equation describing the evolution of $F_{-}^{\rm p}(\zeta_{\rm p})$. It is obtained by replacing $v\to v_0$ in the TF equation Eq. \eqref{TW_eq} specified to $\sigma=-$ and reads
\be
2\frac{v_0}{r_a}\partial_{\zeta_{\rm p}} F_{-}^{\rm p}=-F_{+}^{\rm p}-\kappa F_{-}^{\rm p}+\frac{1+\kappa}{2} F_{-}^{\rm p}\left(F_{-}^{\rm p}+F_{+}^{\rm p}\right)\;,\label{f_m_diff}
\ee
where we remind that $\kappa=-r_s/r_a$. We now use Eq. \eqref{f_p_al} to express $F_{-}^{\rm p}$ in terms of $F_{+}^{\rm p}$, yielding
\be
F_{-}^{\rm p}=\frac{(1+\kappa)F_{+}^{\rm p}-2\kappa}{2-(1+\kappa)F_{+}^{\rm p}}F_{+}^{\rm p}\;.
\ee
Replacing the expression of $F_-^{\rm p}$ in each side of Eq. \eqref{f_m_diff} and after some simplifications, we obtain a first order non-linear differential equation for the TF solution $F_{+}^{\rm p}(\zeta_{\rm p})$,
\be
\frac{v_0}{r_a}\partial_{\zeta_{\rm p}} F_{+}^{\rm p}=-\frac{2(1-\kappa^2) F_{+}^{\rm p}(1-F_{+}^{\rm p})}{4(1+\kappa)F_{+}^{\rm p}-4\kappa-(1+\kappa)^2 {F_{+}^{\rm p}}^2}\;.\label{f_p_diff}
\ee
The function $F_{+}(\zeta_{\rm p})$ is monotonically decreasing in the interval $\zeta_{\rm p}\in (-\infty,0]$ where it takes value from $F_{+}(\zeta_{\rm p}\to -\infty)=1$ to $F_{+}(\zeta_{\rm p}=0)=2\kappa/(1+\kappa)$. One can then define its inverse function $Z_+^{\rm p}(f)$, which is defined in the interval $f\in [2\kappa/(1+\kappa),1]$ and takes values from $Z_+^{\rm p}(f=2\kappa/(1+\kappa))=0$ to $Z_+^{\rm p}(f\to 1)\to -\infty$. Using the relation between the derivatives of inverse functions $\partial_f Z_+^{\rm p}(f)=(\partial_{\zeta_{\rm p}} F_+^{\rm p}(\zeta_{\rm p}))^{-1}$, Eq. \eqref{f_p_diff} yields a differential equation for $Z_+^{\rm p}(f)$ that can be solved exactly as
\begin{align}
\frac{r_a}{v_0}Z_+^{\rm p}(f)&=\int_{\frac{2\kappa}{1+\kappa}}^{f}du\frac{4(1+\kappa) u-4\kappa-(1+\kappa)^2 u^2}{2(1-\kappa^2) u(1-u)}\nn\\
&=\frac{2\kappa}{1-\kappa^2}\ln\left(\frac{1+\kappa}{2\kappa}f\right)-\frac{(1+\kappa)f-2\kappa}{2(1-\kappa)}\nn\\
&+\frac{3+\kappa}{2(1+\kappa)}\ln\left[\frac{1+\kappa}{1-\kappa}(1-f)\right]\;.\label{inv_plus}
\end{align}
Similarly, the function $Z_-^{\rm p}(f)$, inverse of the TF solution $F_-^{\rm p}(\zeta_{\rm p})$ can be obtained by replacing $+\to -$ in the left-hand side of Eq. \eqref{inv_plus} and replacing $f$ by the right-hand-side of Eq. \eqref{Fp_Fm}. Note that this relation has no well-defined limit for $\kappa\to 1$, which is not surprising as in this limit $F_+^{\rm p}(\zeta_{\rm p})=\Theta(-\zeta_{\rm p})$ is not monotonically decreasing and thus does not have an inverse function.
In Fig. \ref{Fig_F_per}, we plot the cumulative probability in the persistent phase obtained by numerical simulation of the stochastic process. The TF contribution shows excellent agreement with our analytical prediction for $F_\sigma^{\rm p}(\zeta_{\rm p})$ obtained by functional inversion of Eq. \eqref{inv_plus} and from the relation \eqref{Fp_Fm}. The definition of the rescaled position  $\zeta_{\rm p}=x-v_0 t$ taken here is such that $F_-(\zeta_{\rm p}=0)$ and thus the front's position is exactly $m(t)=v_0 t$. Taking instead the definition of the front's position $m(t)$ as the position at time $t$ such that $F_\sigma(\zeta_{\rm p}=0)=f$ with $0\leq f<1$, the correction $X(t)=m(t)-v_0 t$ in the large $t$ limit is of order $O(1)$ and depends explicitly on $f$. Its value is exactly given by $X(t)=Z_{\sigma}(f)$ and obtained from Eqs. \eqref{inv_plus} and \eqref{Fp_Fm}. We will now use the exact expression of the inverse function to obtain the asymptotic behaviours of $F_\sigma^{\rm p}(\zeta_{\rm p})$ for $\zeta_{\rm p}\to -\infty$. 

\subsection{Asymptotic behaviour for $\zeta_{\rm p}\to -\infty$}

To obtain this asymptotic behaviour, we use that at leading order, the TF converges to the stable fixed point $F_{\sigma}^{\rm p}(\zeta_{\rm p}\to \infty)\to 1$. We must therefore consider the asymptotic behaviour 
\be
Z_\sigma^{\rm p}(f)\approx \frac{1}{\lambda_{\rm p}}\ln\left(\frac{1-f}{C_{\sigma}}\right)\;,\;\;f\to 1\;,
\ee
obtained from Eq. \eqref{inv_plus}. This yields 
\be
F_{\sigma}^{\rm p}(\zeta_{\rm p}\to -\infty)\approx 1-C_{\sigma}e^{\lambda_{\rm p}\zeta_{\rm p}}\;.\label{as_per_infty}
\ee
Here, the rate $\lambda_{\rm p}$ reads
\be
\lambda_{\rm p}=\frac{r_a}{v_0}\frac{2(1+\kappa)}{3+\kappa}=\frac{1}{v_0}\frac{(2\gamma+1)(1-\delta)}{2(1+\gamma)-\delta}\;,\label{decay_rate_per}
\ee
while the coefficients $C_{\sigma}$ read
\begin{align}
\ln C_{+}=&\frac{4\kappa}{(1-\kappa)(3+\kappa)}\ln\left[\frac{2\kappa}{1+\kappa}\right]+\frac{1+\kappa}{3+\kappa}
+\ln\left[\frac{1-\kappa}{1+\kappa}\right]\,,\nn\\
C_{-}=&\frac{3+\kappa}{1-\kappa} C_{+}\;.
\label{const_per}
\end{align}
The relation between the coefficients $C_+$ and $C_-$ is obtained using the limit $F_{\sigma}^{\rm p}\to 1$ in Eq. \eqref{f_p_al}. In the limit $\kappa\to 1$ with $r_a=1/2$, one recovers that $\lambda_{\rm p}\to (2v_0)^{-1}$ and $C_{-}\to 2$, in full agreement with the asymptotic limit in Eq. \eqref{as_inf_inf_per}. Let us now consider the opposite asymptotic behaviour $\zeta_{\rm p}\to 0$.

\subsection{Asymptotic behaviour for $\zeta_{\rm p}\to 0$}

As we have seen in the previous sections, the solution in the persistent phase has a finite edge for $\zeta_{\rm p}=0$, beyond which it is identically zero. We will now consider the asymptotic behaviour as $\zeta_{\rm p}\to 0$. We suppose that the function $F_{\sigma}^{\rm p}(\zeta_{\rm p})$ is an analytic function close to $\zeta_{\rm p}=0_-$,
\be
F_{\sigma}^{\rm p}(\zeta_{\rm p})=\sum_{k\geq 0} c_{k,\sigma} z^{k}\;,\;\;z=\frac{r_a\zeta_{\rm p}}{v_0}\;,\label{gen_Tayl}
\ee
with at leading order $c_{0,-}=F_-^{\rm p}(0)=0$ and $c_{0,+}=F_+^{\rm p}(0)=2\kappa/(1+\kappa)$. Note that this hypothesis holds for $\sigma=-$ in the special case of infinite persistence ($\kappa\to 1$). Inserting directly in Eqs. \eqref{f_m_diff} and \eqref{f_p_al} the Taylor series \eqref{gen_Tayl} for the TF, we can compute recursively the Taylor coefficients
\begin{align}
2 k\, c_{k+1,-}=&-c_{k,+}+\kappa c_{k,-}\label{Tayl_m}\\
&+\frac{1+\kappa}{2}\sum_{p=0}^k c_{k-p,-}(c_{p,+}+c_{p,-})\;,\nn\\
\kappa c_{k,+}+c_{k,-}&=\frac{1+\kappa}{2}\sum_{p=0}^k c_{k-p,+}(c_{p,+}+c_{p,-})\;.\label{Tayl_p}
\end{align}
Computing these coefficients up to order $z^2$, the asymptotic behaviours as $\zeta_{\rm p}\to 0$ read
\begin{align}
F_-^{\rm p}(\zeta)&=-\frac{\kappa}{1+\kappa}z+\frac{1-\kappa}{4(1+\kappa)}z^2+O(z^3)\;,\label{Fm_sm}\\
F_+^{\rm p}(\zeta)&=\frac{2\kappa}{1+\kappa}-\frac{1-\kappa}{1+\kappa}z-\frac{1+\kappa}{4\kappa}z^2+O(z^3)\label{Fp_sm}\;,
\end{align}
where we remind that $z=r_a\zeta_{\rm p}/v_0$ and $\kappa=-r_s/r_a$. In the case of infinite persistence, taking the limit $\kappa\to 1$ and using that $r_a=1/2$, one can check that the asymptotic behaviour in Eq. \eqref{Fm_sm} is in full agreement with Eq. \eqref{as_0_inf_per}. We will now complete our analysis of the persistent phase by computing the moments of the cumulative distribution described by the TF $F_{\sigma}(\zeta_{\rm p})$.

\subsection{Moments of the distribution}

\begin{figure}
    \centering
    \includegraphics[width=0.475\textwidth]{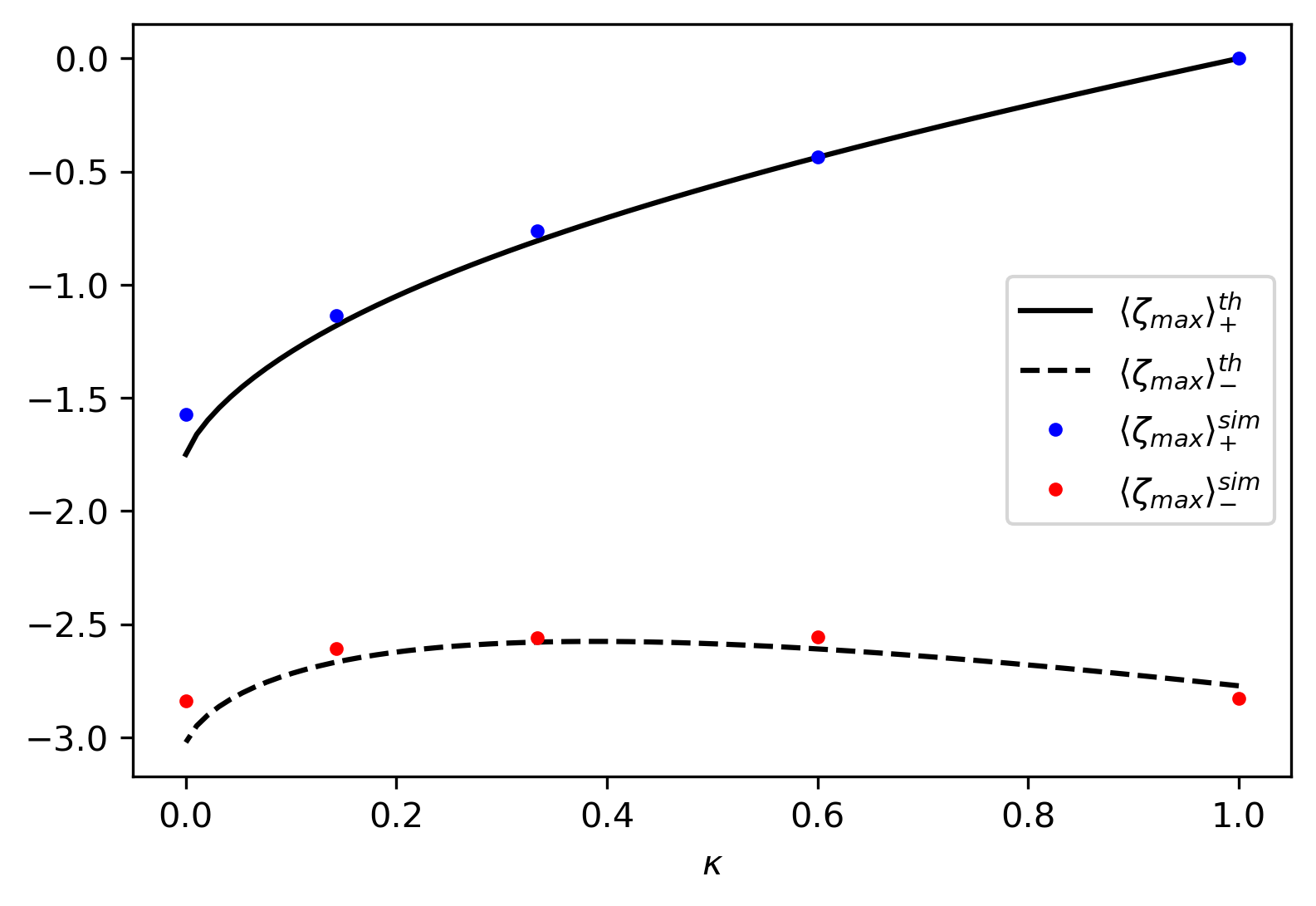}
    \caption{Plot of the average value $\moy{\zeta_{\max}}_{\sigma}=\moy{x_{\max}(t)-v_0 t}_{\sigma}$ obtained by numerical simulation of the stochastic process starting with direction $\sigma=\pm$ for $\delta=0$ and $\gamma=0,0.125,0.25,0.375,0.5$ in the persistent phase ($\gamma<1/2$) as a function of $\kappa=(1-2\gamma)/(1+2\gamma)$ (taking $v_0=1$). The numerical results show excellent agreement with the analytical prediction $\moy{\zeta_{\max}}_{\pm}=-(r_a/v_0){\cal M}_{\pm}(\kappa)$, where the scaling functions ${\cal M}_{\pm}(\kappa)$ are given in Eqs. \eqref{M_+} and \eqref{M_-} and $r_a=1/(1+\kappa)$.
    }
    \label{Fig_mean}
\end{figure}

In the intermittent phase, we have obtained the average value for the maximum in the large time limit $\moy{x_{\max}(t)}_{\sigma}=v_{\rm i} t+3/(2\lambda_{\rm i})\ln t+O(1)$, starting the process in direction $\sigma=\pm$, up to corrections of $O(1)$. In the persistent phase instead, we will now see that using our exact result for the inverse function $Z_{+}^{\rm p}(f)$ it is possible to obtain the exact value of the moments of $x_{\max}(t)$. More precisely, conditioning on the survival of the process (or without conditioning if $\delta=0$), the rescaled random variable
\be
\zeta_{\max}=x_{\max}(t)-v_0 t\;,
\ee
becomes independent of $t$ in the large time limit and its cumulative distribution is $F_{\sigma}^{\rm p}(\zeta_{\rm p})$. Taking the definition of the front's position as
$m(t)=\moy{x_{\max}(t)}_{\sigma}$ conditioned on the survival of the process ($N(t)>0$), the average $\moy{\zeta_{\max}}_{\sigma}=X(t)=m(t)-v_0 t$ gives the $O(1)$ correction in the large time limit. The moments of this random variable read
\be
\moy{\zeta_{\max}^n}_{\sigma}=-\int_{-\infty}^{0}\partial_{\zeta}F_\sigma^{\rm p}(\zeta) \zeta^n d\zeta\;.\label{z_max_mom}
\ee
Introducing a change of variable from $\zeta\to f=Z_{\sigma}^{\rm p}(\zeta)$, one obtains the following representation for the moments
\be
\moy{\zeta_{\max}^n}_{\sigma}=\int_{F_{\sigma}^{\rm p}(0)}^{1}\left[Z_{\sigma}^{\rm p}(f)\right]^n df\;,\label{moy_inv}
\ee
where we remind that $F_{-}^{\rm p}(0)=0$ while $F_{+}^{\rm p}(0)=2\kappa/(1+\kappa)$.
Using the expression of the inverse function in Eq. \eqref{inv_plus}, the average value can be computed analytically and takes the scaling form
\begin{align}
\moy{\zeta_{\max}}_{\pm}&=-\frac{v_0}{r_a}{\cal M}_\pm\left(\kappa\right)\;,\label{mean}\\
{\cal M}_+(\kappa)&=\frac{7-3\kappa}{4(1+\kappa)}+\frac{2\kappa}{1-\kappa^2}\ln\left(\frac{2\kappa}{1+\kappa}\right)\;.\label{M_+}\\
{\cal M}_-(\kappa)&=\frac{1+3\kappa+16\ln 2}{4(1+\kappa)}+\frac{2\kappa}{1-\kappa^2}\ln\left(\frac{2\kappa}{1+\kappa}\right)\;.\label{M_-}
\end{align}
 Taking the limit $\kappa\to 1$ and $r_a=1/2$ in Eq. \eqref{mean}, we obtain $\moy{\zeta_{\max}}_{+}=0$ and $\moy{\zeta_{\max}}_{-}=-4v_0 \ln2$, which coincides with the result for infinite persistence, taking the limit $n\to 1$ of Eq. \eqref{mom_triv}. In Fig. \ref{Fig_mean}, we compare the analytical prediction for $\moy{\zeta_{\max}}_{\pm}$ given by the scaling form in Eq. \eqref{mean} and the scaling functions in Eqs. \eqref{M_+} and \eqref{M_-} to the value obtained by numerical simulation of the process for $\delta=0$, showing good agreement. The agreement is poorer for $\kappa=0$, corresponding to the transition. We are now going to detail this case and see in particular that this behaviour is due to the finite time effects ($t=20$ for the data on the graph), which are particularly strong at the transition.


\section{Transition}\label{sec_ph_trans}

We have characterised in section \ref{sec_int_phase} the TF in the intermittent phase, for $r_s=\delta+\gamma-1/2>0$ and in the persistent phase, i.e. $r_s<0$ in section \ref{sec_per_phase}. We will now discuss its properties at the transition, i.e. for $r_s=0$, where it shares features of both phases.

\subsection{Solution in the large time limit}

The velocity of the travelling front at the transition is $v_0$ as in the persistent phase. However we will show that in the large time limit, there is no discontinuity of the distribution starting in the state $\sigma=+$, i.e. $F_{+}^{\rm t}(\zeta_{\rm p}=x-v_0 t=0)=0$, with the superscript $^{\rm t}$ corresponding to the transition. There is a simple relationship between the TWs $F_+^{\rm t}(\zeta_{\rm p})$ and $F_-^{\rm t}(\zeta_{\rm p})$ at the transition, obtained by taking the limit $\kappa\to 0$ in Eq. \eqref{Fp_Fm}, 
\be
F_+^{\rm t}=-\frac{F_-^{\rm t}}{2}+\frac{1}{2}\sqrt{F_-^{\rm t}(8+F_-^{\rm t})}\;.\label{z_t_m}
\ee
We can also obtain an exact expression for the inverse function $Z_{+}^{\rm t}(f)$ of the TF solution $F_+^{\rm t}(\zeta_{\rm p})$ by taking the limit $\kappa\to 0$ in Eq. \eqref{inv_plus}, yielding
\be
\frac{r_a}{v_0}Z_{+}^{\rm t}(f)=\frac{3}{2}\ln(1-f)-\frac{f}{2}\;.\label{z_t_p}
\ee
Using this expression in the limit $f\to 1$, one can show that the solution behaves asymptotically as
\be
F_\sigma^{\rm t}(\zeta_{\rm p})\approx 1-C_{\sigma}^{\rm t}e^{\frac{2r_a\zeta_{\rm p} }{3v_0}}\,,\;C_{+}^{\rm t}=e^{-\frac{1}{3}}\,,\;C_{-}^{\rm t}=3e^{-\frac{1}{3}}\,,\label{F_t_inf}
\ee
in full agreement with the limit $\kappa=-r_s/r_a\to 0$ of Eqs. \eqref{decay_rate_per} and \eqref{const_per}. However, the asymptotic behaviour of the distribution $F_{\sigma}^{\rm t}(\zeta_{\rm p}\to 0)$ is not correctly reproduced by Eqs. \eqref{Fm_sm} and \eqref{Fp_sm}. One needs instead to take first the limit $\kappa\to 0$ in Eqs. \eqref{Tayl_m} and \eqref{Tayl_p}, and then solve recursively for the Taylor coefficients. It yields
\begin{align}
F_-^{\rm t}(\zeta_{\rm p})&=\frac{z^2}{8}+\frac{z^3}{64}-\frac{7}{2048}z^4+O(z^5)\;,\label{Fm_t_sm}\\
F_+^{\rm t}(\zeta_{\rm p})&=-z-\frac{3}{32}z^2-\frac{z^3}{256}+\frac{9}{8192}z^4+O(z^5)\label{Fp_t_sm}\;,
\end{align}
where we remind that $z=r_a\zeta_{\rm p}/v_0$. 
Finally, one can check that replacing $Z_{+}^{\rm p}(f)\to Z_{+}^{\rm t}(f)$ in the equation for the moments Eq. \eqref{moy_inv} that the first moments $\moy{\zeta_{\max}}_{\pm}$ are correctly reproduced by Eq. \eqref{mean} by taking the limit $\kappa\to 0$ of the scaling functions ${\cal M}_{\pm}(\kappa)$ with ${\cal M}_+(0)=7/4$ and ${\cal M}_-(0)=1/4+4\ln 2$. 
\begin{figure}
    \centering
    \includegraphics[width=0.475\textwidth]{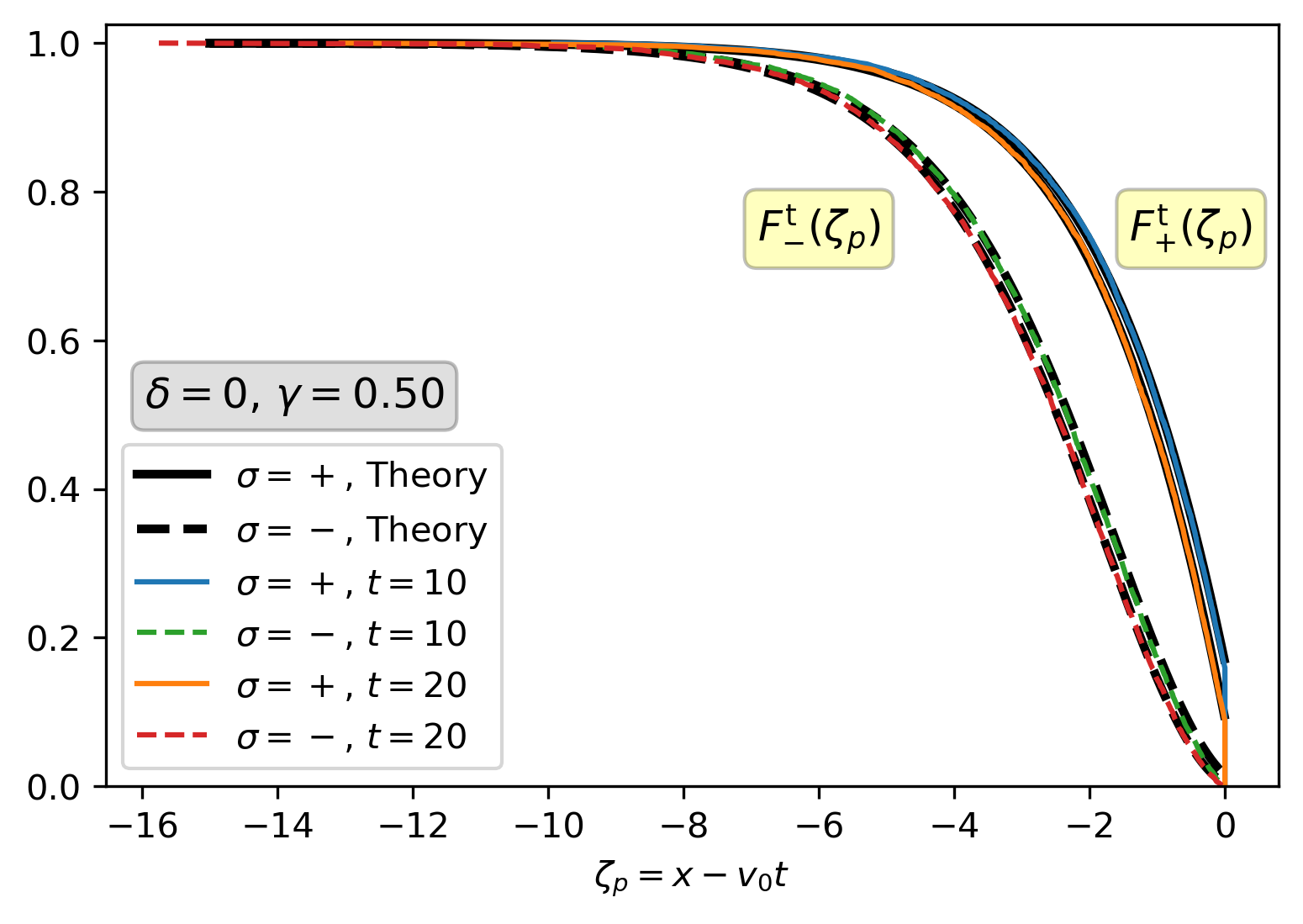}
    \caption{Plot of $Q_{\pm}(x,t)$ versus the rescaled position $\zeta_{\rm p}=x-v_0 t$ obtained from numerical simulations of the stochastic process at the transition ($\gamma+\delta=1/2$). We chose $\delta=0$ in this figure such that there is no stationary state contribution to $Q_{\pm}(x,t)$. The numerical results for the TF solution with initial direction $\sigma=\pm$ and for $t=10,20$ show a good collapse on the analytical results $F_{\pm}^{\rm t}(\zeta_{\rm p})$ obtained by functional inversion of Eq. \eqref{z_t_p_finite_t} (with respect to $f$) at the corresponding times and using Eq. \eqref{z_t_m}.
    }
    \label{Fig_F_trans}
\end{figure}

\subsection{Finite time effects}

In Fig. \ref{Fig_F_trans}, we plot the TF solution obtained by numerical simulation of the stochastic process at the transition in the case where $\delta=0$ and $\gamma=1/2$.
In this case, the cumulative distribution at time $t\gg 1$ is completely described by the TF, $Q_{\sigma}(x,t)=F_{\sigma}(\zeta_{\rm p}=x-v_0 t)$. For smaller times of order $t=O(1)$, the cumulative probability depends both on $x$ and $t$ in a separate manner and not only on $\zeta_{\rm p}=x-v_0 t$. Note that the TWs $F_{+}^{\rm t}(\zeta_{\rm p})$ in Fig. \ref{Fig_F_trans} exhibits for $\zeta_{\rm p}=0$ a discontinuity at the edge for finite time. 
This discontinuity decreases with time and cannot be obtained from Eq. \eqref{z_t_p} which corresponds to the limit $t\to \infty$, for which the discontinuity has vanished. As we have already seen for the persistent phase in section \ref{sec_per_phase}, the exact value of $Q_+(v_0 t,t)$ is given by the probability that the cluster initiated from the initial particle contains a non-zero number of particles at time $t$. At the transition, the branching rate within a cluster and the rate at which particles separate from this cluster are both equal to $1/2$. Using the results of App. \ref{app_num} and in the limit of equal rates, we can compute the probability 
\be
Q_+(v_0 t,t)=\Prob\left[x_{\max}(t)=v_0 t|\sigma=+\right]=\frac{2}{2+t}\;.
\ee
At the transition, this probability converges algebraically to zero whereas both in the intermittent and persistent phase, the convergence to zero and $1-2\gamma-2\delta$ respectively is exponential. For sufficiently large time, we expect the TF evolution equation in Eq. \eqref{f_p_diff} to hold. Using this boundary condition for finite time, we obtain the finite time corrected function $Z_{+}^{\rm t}(f;t)$, inverse of the TF $F_+^{\rm t}(\zeta_{\rm p})$,
\be
\frac{r_a}{v_0}Z_{+}^{\rm t}(f;t)=\frac{3}{2}\ln\left(\frac{2+t}{t}(1-f)\right)-\frac{f}{2}+\frac{1}{2+t}\;.\label{z_t_p_finite_t}
\ee
Using additionally the relation between the TF solutions in Eq. \eqref{z_t_m}, we compare in Fig. \ref{Fig_F_trans} our finite time approximation to the results from numerical simulation. For the times $t=10,20$, this analytical result gives an excellent approximation to the solution. Note that in the general case where $\delta>0$, the relation in Eq. \eqref{late_t_cond} between the full CDF $Q_{\sigma}(x,t)$ and the TF $F_{\sigma}(\zeta_{\rm p})$ is subject to additional finite time corrections such that Eq. \eqref{z_t_p_finite_t} is only valid for $t\gg 1$.

\section{Conclusion}\label{sec_con}

In this paper, we have analysed analytically the extreme value statistics (EVS) of 
a  model of branching run-and-tumble particles in one dimension. In the large time limit, the cumulative distribution of the maximum of the process $x_{\max}(t)$ is described by a travelling front (TF). We have recovered the exact results for the velocity and obtained the corrections to the front's position and the shape of the TF. 
This model exhibits a phase transition between a persistent and an intermittent phase \cite{demaerel2019asymptotic,HORSTHEMKE1999285} which arises from respectively the presence or absence of long-lived macroscopic clusters of particles and that is strongly reflected in its EVS. While the TF solution in the intermittent phase shares a number of qualitative features with the TF describing the maximum of a BBM, the TF solution in the persistent phase exhibits completely novel features. In particular it exhibits a finite edge beyond which it vanishes exactly even in the limit $t\to \infty$ and is discontinuous at this edge. This discrepancy can be traced back to the persistence of the underlying branching random walk, seen in RTP but absent for Brownian motion. This toy model does not reproduce all the features of a spreading colony of bacteria and we expect that some characteristics described in the persistent phase are not robust and model dependent. However, it has the large advantage to be fully analytically tractable, allowing for a good understanding of the underlying physical properties of the process. Note also that it constitutes one of the rare models for which a full analytical description of the travelling front solution for the EVS is possible.

We focused here only on the TF solution describing the EVS conditioned on the survival of the process, i.e. a number $N(t)>0$ of particles at time $t$. One could derive instead the stationary distribution $Q_{\sigma}^{\rm st}\left(x\right)$ reached by $x_{\max}$ once all the particles have died. It is particularly important for $\delta\geq 1$, where all particles die eventually in the process and no TF solution exists. As seen from Figs. \ref{Fig_F_int} and \ref{Fig_F_per}, the distribution $Q_{-}^{\rm st}\left(x\right)$ also exhibits a discontinuity for $x=0$ that would be interesting to characterise.

Another interesting quantity associated to the EVS is the time $t_{\max}$ at which the maximum $x_{\max}(t)$ of the process up to time $t$ is reached. In the persistent phase, as $x_{\max}(t)$ grows at all time with speed $v_0$, one can conjecture that $t_{\max}=t$. These problems are left for future investigation.

{\bf Acknowledgement}: We would like to thank D. Mukamel and O. Raz for critically reading this manuscript and for their helpful remarks. We also thank S.N. Majumdar and G. Schehr for their interesting comments and for pointing out useful references.

\appendix

\section{Evolution of the number of particles} \label{app_num}

In order for this article to be self-contained, we detail a few properties on the statistics of the number of particles in branching processes.
We consider a branching process with branching rate $b$ and dying rate $d<b$. Initially, there is only one particle in the system and we want to obtain the probability for the number $N(t)$ of particles at time $t$. We denote $P_{n}(t)=\Prob\left[N(t)=n\right]$ the probability that there are $n$ particles in the system at time $t$. The evolution of this probability in the small interval $dt$ can be derived using a backward Fokker-Planck equation \cite{ramola2015branching}. We consider all evolution which lead to a number $N(t)=n$ of particles at time $t$. The number of particle stays the same with probability $1-(b+d)dt$ during the initial interval. It goes from $1$ to $2$ with probability $b dt$. Afterwards the two particles evolve independently and can give rise to an arbitrary number $m$ and $n-m$ of particles. Finally, if the initial particle dies (which happens with probability $d dt$), the system can only have zero particle at time $t$.
This yields the differential equation
\be
\partial_t P_n(t)=-(b+d)P_n(t)+b\sum_{m=0}^n P_{m}(t)P_{n-m}(t)+d\delta_{n,0}\;.
\ee
Introducing the generating function $P(s;t)=\sum_{n\geq 0} s^n P_n(t)$, one obtains the equation
\be
\partial_t P(s;t)=(1-P(s;t))(d-b P(s;t))\;,
\ee
with the initial condition $P(s;0)=s$. The generating function is obtained exactly as
\be\label{GF_num}
P(s;t)=\frac{d(1-s)e^{(b-d)t}-(d- b s)}{b(1-s)e^{(b-d)t}-(d- b s)}\;.
\ee
Taking $s=0$, one obtains immediately that for $b>d$,
\be
P_0(t)=\Prob\left[N(t)=0\right]=\frac{d(e^{(b-d)t}-1)}{b e^{(b-d)t}-d}\to \frac{d}{b}\;.
\ee
The probability that the process has survived up to time $t$ simply reads $P(t)=\Prob\left[N(t)>0\right]=1-P_0(t)$. In the limit where $b=d$, the probability simplifies to 
\be
P_0(t)=\Prob\left[N(t)=0\right]=\frac{bt}{1+bt}\;.
\ee
More generally, the probability that there are $n$ particles at time $t$ reads
\be
P_n(t)=\left(\frac{b-d}{b}\right)^2\frac{\left(1-e^{-(b-d)t}\right)^{n-1}}{\left(1-\frac{d}{b}e^{-(b-d)t}\right)^{n+1}}e^{-(b-d)t}\;.
\ee
In the special case where $d=0$, this expression simplifies to
\be
P_n(t)=\left(1-e^{-bt}\right)^{n-1}e^{-b t}
\ee
Finally, the average number of particles at time $t$ can also be obtained from Eq. \eqref{GF_num} as
\be
\moy{N(t)}=\left.\partial_s P(s;t)\right|_{s=1}=e^{(b-d)t}\;.
\ee



\bibliography{bRTP}

\end{document}